\documentclass{aa}  

\usepackage{graphicx}
\usepackage{epsf,epsfig,amsmath,amssymb,pifont,rotate}
\usepackage{txfonts}
\usepackage{natbib}

    \def\half{{\scriptstyle{}^1\!\!/\!{}_2}}

    \def\etal{${\rm \hspace*{0.7ex}et\hspace*{0.7ex}al.\hspace*{0.6ex}}$}
    \def\plus{${\rm \hspace*{0.6ex}\&\hspace*{0.6ex}}$}
    \def\ie{i.\,e.\ }
    \def\eg{e.\,g.\ }

    \def\vdreq{v^{\hspace{-0.8ex}^{\circ}}_{\rm dr}}

    \def\pabl#1#2{\frac{{\rm\partial} #1}{{\rm\partial} #2}}
    \def\Vl{{V_{\ell}}}
    \def\Vs{{V_{\rm s}}}
    \def\xx{\vec{x}}

    \def\nH{n_{\langle{\rm H}\rangle}}
    \def\tmix{\tau_{\rm mix}} 
    \def\Nl{{N_{\ell}}}
    \def\Sr{S_{\!r}}

\begin{document}
   \title{Dust in Brown Dwarfs and Extra-solar Planets}

   \subtitle{I. Chemical composition and spectral appearance 
             of quasi-static cloud layers}

   \author{Ch. Helling\inst{1}
          \and
           P. Woitke\inst{2}
          \and
           W.-F. Thi\inst{3}
          }

   \offprints{Ch. Helling (Christiane.Helling@st-andrews.ac.uk)}

   \institute{SUPA, School of Physics \& Astronomy, University of St Andrews, 
              North Haugh, St Andrews,  KY16 9SS, Scotland, UK\\
              \email{Christiane.Helling@st-andrews.ac.uk}
   \and
              UK Astronomy Technology Centre, Royal Observatory, Blackford Hill, Edinburgh EH9 3HJ, 
              Scotland, UK
   \and
            SUPA, Institute for Astronomy, The University of Edinburgh, Royal Observatory, Blackford Hill, Edinburgh
EH9 3HJ, Scotland, UK
             }

  \date{July 2007; March 2008}

  \abstract 
%% context heading (optional)
 {}
%% Aim (mandatory) 
  {{Brown dwarfs are covered by dust cloud layers which cause
  inhomogeneous surface features and move below the observable
  $\tau=1$ level during the object's evolution.  
  In any case, the
  cloud layers have a strong influence on the structure and spectral
  appearance of brown dwarfs and extra-solar planets, \eg by providing
  large local opacities and by removing condensable elements from the
  atmosphere causing a sub-solar metalicity in the atmosphere. 
  We aim
  at understanding the formation of cloud layers in quasi-static
  substellar atmospheres which consist of dirty grains composed of
  numerous small islands of different solid condensates.}}
%% methods heading (mandatory) 
  {The time-dependent description presented in (Helling \& Woitke
  2006) is a kinetic model describing nucleation, growth and
  evaporation. It is extended to treat gravitational settling and is
  applied to the static-stationary case of substellar model
  atmospheres.  From the solution for the dust moments, we determine
  the grain size distribution function approximately which, together
  with the calculated material volume fractions, provides the basis
 for applying effective medium theory and Mie theory
  to calculate the  opacities of the composite dust grains.}
%% results heading (mandatory)
  {The cloud particles in brown dwarfs and hot giant-gas planets
  are found to be small in the high atmospheric layers
  ($a\!\approx\!0.01\,\mu$m), and composed of a rich mixture of all
  considered condensates, in particular the abundant MgSiO$_3$[s],
  Mg$_2$SiO$_4$[s] and SiO$_2$[s]. As the particles settle downward,
  they increase in size and reach several $100\,\mu$m in the deepest
  layers. The more volatile parts of the grains evaporate and the
  particles stepwise purify to form composite particles of
  high-temperature condensates in the deeper layers, mainly Fe[s] and
  Al$_2$O$_3$[s]. The gas phase abundances of the elements involved in
  the dust formation process vary by orders of magnitudes
  throughout the atmosphere. The grain size distribution is found to
  be relatively broad in the upper atmospheric layers but strongly
  peaked in the deeper layers. This reflects the cessation of the
  nucleation process at intermediate heights. The spectral appearance
  of the cloud layers in the mid IR ($7-20\,\mu$m) is close to a grey
  body with only weak broad features on a few percent level, mainly
  caused by MgSiO$_3$[s], and Mg$_2$SiO$_4$[s]. These features are,
  nevertheless, a fingerprint of the dust in the higher atmospheric
  layers that can be probed by observations.}
%% conclusions heading (optional)
  {Our models predict that the gas phase depletion is much  weaker
  as compared to phase-equilibrium calculations in the high
  atmospheric layers. Because of the low densities, the dust formation
  process is incomplete there, which results in considerable amounts
  of left-over elements that might produce stronger and broader
  neutral metallic lines. }


  \maketitle
%
%________________________________________________________________

\section{Introduction} 

Dust in form of small solid particles ({\it grains}) becomes an
increasingly important component in understanding the nature of
substellar objects with decreasing $T_{\rm eff}$, i.e. brown dwarfs
and giant-gas planets. Observations start to provide direct evidence
for dust clouds covering brown dwarfs (Cushing et al. 2006) and
extrasolar giant-gas planets (Richardson et al. 2007, Swain et
al. 2007).  The search for biosignatures in extraterrestrial planets
becomes more complicated if such cloud layers cover the atmosphere
that efficiently absorb in the wavelength region where e.g. the Earth
vegetation's red edge spectroscopic features ($600-1100\,$nm, Saeger
et al. 2007) or the extraterrestrial equivalents are situated.  In
fact, the development of live seems impossible below optically
thick cloud layers, because the star light needs to reach the surface
to create the necessary departures from thermodynamical equilibrium
that allow for structure formation.  Furthermore, abundances of
molecules like O$_2$ and O$_3$ as the carriers of spectral
biosignatures will be strongly affected by the presence of cloud
layers, because of chemical surface reactions and the element
depletion due to dust formation in the atmosphere.  Thus, the
understanding of the details of cloud formation physics and chemistry
is a major issue in modelling substellar atmospheres.

This paper presents a kinetic approach for modelling quasi-static
atmospheres with stationary dust cloud layers including seed particle
formation (nucleation), grain growth, gravitational settling, grain
evaporation, and element conservation (Sects.~\ref{sec:dustmodel} and
\ref{sec:mat}). The model is a further development of the
time-dependent description presented in Helling \& Woitke (2006;
Paper~V) and is particularly suited to treat the formation of
``dirty'' dust grains (\ie particles composed of numerous small
islands of different solid condensates) in the frame of classical
stellar atmospheres with consistent radiative transfer and convection
(Dehn 2007, Helling et al. 2008). The results of the models (see
Sect.~\ref{sec:results}) describe the vertical cloud structure and the
amount of dust formed in the atmosphere of substellar objects as well
as the amount of condensable elements left in the gas phase. The
models provide further details like the material composition of the
cloud particles, the mean grain sizes and the size distributions as
function of atmospheric height.  Section~\ref{sec:srt} demonstrates
what spectral features such cloud layers made of dirty solid particles
exhibit, and from which temperature and pressure level these features
originate.

%__________________________________________________________________

\section{Nucleation, growth and evaporation of precipitating 
         dirty grains}
\label{sec:dustmodel}

In Paper~V, a kinetic description for the nucleation, growth and
evaporation of dirty dust particles has been developed by extending
the time-dependent moment method of Gail\plus Sedlmayr (1988). The
basic idea is that a ``dirty'' solid mantle will grow on top of the
seed particles, because these seeds can only form at relatively low
temperates ($T\!\la\!1400\,$K) where the oxygen-rich gas is strongly
supersaturated with respect to several solid materials. The dirty
mantle is assumed to be composed of numerous small islands of
different pure condensates. The formation of islands is motivated by
experiments in solid state physics (Ledieu\etal 2005, also
Zinke-Allmang 1999) and by observations of coated terrestrial dust
particles (Levin\etal 1996, Korhonen\etal 2003).  Note that we
consider dust formation by gas-solid reactions only and omit
solid-solid reactions and lattic rearrangements inside the
grains in our model.

In the following, we extent this description to include the effects of
{\it gravitational settling} ({\it drift, rain-out, precipitation})
which is important to understand the long-term quasi-static structures
of brown dwarf and gas-giant atmospheres. The challenge here is to account properly
for the element conservation when dust particles consume certain
elements from the gas phase at the sites of their formation, transport
them via drift motions through the gas and finally release the
elements by evaporation at other places. Furthermore, we want to
abandon the assumption that the number of elements equals the number
of condensates made in Paper~V, because there are typically much more
condensates than elements.

We consider the moments $L_{\rm j}(\xx,t)$ [cm$^j$g$^{-1}$]
($j\!=\!0,1,2,...$) of the dust volume distribution function
$f(V,\xx,t)$ [cm$^{-6}$] (for more details, see Paper~V, Sect. 2.1),
where $\xx$ and $t$ are space and time. $V$ [cm$^3$] is the volume of
an individual dust particle. The total dust volume per cm$^3$ stellar
matter, $V_{\rm tot}$, is given by the 3$^{\rm rd}$ dust moment as
\begin{equation} 
\label{eq:Vtot} 
  \rho\,L_3 = \int_{V_\ell}^{\infty}\!\!f(V)\,V\,dV 
       \;\;=\;\; V_{\rm tot}\quad[\rm cm^3\ cm^{-3}]\ ,
\end{equation}
where $\rho$ [g\,cm$^{-3}$] is the mass density and $V_{\ell}$ is the lower
integration boundary. In an analogous way, we define the volume $\Vs$
of a certain solid species $s$ by
\begin{equation} 
\label{eq:Vs1}
  \rho\,L_3^{\rm s} = \int_{V_\ell}^{\infty}\!\!f(V)\,V^{\rm\,s}\,dV 
       \;\;=\;\; \Vs\quad[\rm cm^3\ cm^{-3}]\ ,
\end{equation}
where $V^{\rm\,s}$ [cm$^3$] is the sum of island volumes of material s
in {\it one} individual dust particle.  For simplicity, we assume that
$V^{\rm\,s}/V\!=\!\Vs/V_{\rm tot}$ is constant for all dust particles
at a certain position in the atmosphere, \ie we assume a unique
volume composition of all grains at $(\xx,t)$, such that $V_{\rm
tot}\!=\!\sum \Vs$ and $L_3\!=\!\sum L_3^{\rm s}$.

By means of this assumption, it is possible to express the integrals
that occur after integrating the master equation (Eq.\,(1) in Paper V)
over size in terms of other moments. The results are the dust moment
conservation equations.  The change of the partial volume of
solid $s$  can then be expressed analog to Eq.\,(23) in Paper~V
\begin{eqnarray}
\label{eq:coneq}
  \pabl{}{t}\left(\rho L_3^{\rm s}\right) 
  \,+\, \nabla\,\Bigg(\int_{V_\ell}^{\infty}\!\!f(V)\,V^{\rm\,s}
                  \!\left(\mathbf{v}_{\rm gas} 
                  \!+\mathbf{\vdreq}(V)\right) dV\Bigg) 
  \nonumber\\
  = \Vl^{\rm s} J_\star + \chi_{\rm net}^{\rm\,s}\,\rho L_2 \,
\end{eqnarray}
where we have assumed that the Knudsen numbers are large (${\rm
Kn}\!\gg\!1$, compare Paper~II) and that the drift velocities
$\mathbf{v}_{\rm dr}(V)$ of the dust particles can be approximated by
the equilibrium drift velocities $\mathbf{\vdreq}(V)$ (final
fall speeds).

The source terms on the r.h.s. of Eq.\,(\ref{eq:coneq}) describe the
effects of nucleation, growth and evaporation of condensate s.
$J_\star\!=\!J(\Vl)\!=\!f(\Vl)\,\frac{dV}{dt}\big\vert_{V=V_l}$
[s$^{-1}$cm$^{-3}$] is the stationary nucleation rate (see
Sect.~\ref{sec:nuc}). $\Vl^{\rm s}$ [cm$^3$] is the volume occupied
by condensate $s$ in the seed particles when they enter the integration
domain in size space. The net growth velocity of condensate s
$\chi_{\rm net}^{\rm\,s}$ [cm\,s$^{-1}$] (negative for evaporation) is given
by (Eq.(24) in Paper V)
\begin{equation}
\chi_{\rm net}^{\rm\,s}
  = \sqrt[3]{36\pi}\,\sum^R_{r=1} 
    \frac{\Delta V_r^{\rm\,s} n_r^{\rm key} {\rm v}_r^{\rm rel} \alpha_r}
         {\nu_r^{\rm key}}
    \left(1 - \frac{1}{\Sr\,b^{\rm\,s}_{\rm surf}}\right) \ .
\label{eq:chinet}
\end{equation}
Here, $r$ is an index for the chemical surface reactions (see
Table~\ref{tab:chemreak}), $\Delta V_r^{\rm\,s}$ is the volume
increment of solid $s$ by reaction $r$ ($\sum \Delta
V_r^{\rm\,s}\!=\!\Delta V_{\rm r}$), $n_r^{\rm key}$ is the particle
density of the key reactant, ${\rm v}_r^{\rm rel}$ is its thermal
relative velocity and $\alpha_r$ is the sticking coefficient of
reaction $r$. $\Sr$ is the reaction supersaturation ratio and
$b^{\rm\,s}_{\rm surf}\!=\!V_{\rm tot}/V_{\rm s}$ is a $b$-factor 
which describes the probability to find a surface of kind $s$ on the
total surface. Putting $b^{\rm\,s}_{\rm surf}$ independent of $V$, we
assume that all grains at a certain point in the atmosphere have the
same surface and volume composition, i.e. the grain material is a
homogeneous mix of islands of different kinds (for more details see
Paper V).

The divergence of the drift term in Eq.\,(\ref{eq:coneq}) is treated
in the following way. Assuming large Knudsen numbers and subsonic
drift velocities, the equilibrium drift velocity is given by
$\mathbf{\vdreq}(V)\!=\!-\vec{e_z}\,a\sqrt{\pi}\,g\,\rho_{\rm
d} /(2\,\rho\,c_{\rm T})$ (Eq.\,63 in Paper~II), where $a$ is the
particle radius, $\vec{e_z}$ the vertical unit vector (pointing
upwards) and $g$ the gravitational acceleration
(downwards). $\rho_{\rm d}\!=\!\sum \rho_{\rm s}\,\Vs/V_{\rm tot}$ is
the dirty dust material density and $\rho_{\rm s}$ the material
density of a pure condensate s. $c_T\!=\!\sqrt{2kT/\bar{\mu}}$ is a
mean thermal velocity with $T$ being the temperature, $k$ the Boltzmann
constant and $\bar{\mu}$ the mean molecular weight of the gas
particles. Inserting this formula into the drift term in
Eq.\,(\ref{eq:coneq}) yields
\begin{equation}
  \int_{V_\ell}^{\infty}\!\!f(V)\,V^{\rm\,s}\,
     \mathbf{\vdreq}(V)\,dV 
  \,=\, -\,\vec{e_z}\,\xi_{\rm lKn}\,\frac{\rho_{\rm d}}{c_T}\,
     \frac{\Vs}{V_{\rm tot}} L_4
\end{equation}
with the abbreviation $\xi_{\rm
lKn}\!=\!\big(\frac{3}{4\pi}\big)^{1/3} \frac{\sqrt{\pi}}{2} g$ (note
the difference to Eq.\,(3) in Paper~III).  Defining $L_4^{\rm
s}\!=\!L_4\,\Vs/V_{\rm tot}$, Eq.\,(\ref{eq:coneq}) in this paper can
be rewritten as
\begin{equation}
\label{eq:Vsequ}
  \pabl{}{t}\big(\rho L_3^{\rm s}\big) 
  + \nabla\,\big(\rho L_3^{\rm s}\,\mathbf{v}_{\rm gas}\big) 
  = \Vl^{\rm s} J_\star + \chi_{\rm net}^{\rm\,s}\,\rho L_2\, 
  +\,\xi_{\rm lKn} \pabl{}{z} \left(\frac{\rho_{\rm d}}{c_T} 
                  L_4^{\rm s}\right)
\end{equation}
Equation~(\ref{eq:Vsequ}) describes the evolution of the partial dust
volume of solid $s$ in space and time due to advection, nucleation,
growth, evaporation {\it and} drift.

The third moment equation for the total dust volume $\rho L_3$
(Eq.\,(1) in Paper~III) can be retrieved by summing up the
contributions from all condensates $s$ as given by
Eq.\,(\ref{eq:Vsequ}), because
\begin{equation}
\sum_{\rm s} L_3^{\rm s} \!=\! L_3 \;,\;
\sum_{\rm s} L_4^{\rm s} \!=\! L_4 \;,\;
\sum_{\rm s} \chi_{\rm net}^{\rm\,s} \!=\! \chi^{\rm net} \;,\;
\sum_{\rm s} \Vl^{\rm s} \!=\! \Vl\ . 
\end{equation}

\subsection{Quasi-static, stationary case}

In the case of a plane-parallel quasi-static stellar atmosphere
$\mathbf{v_{\rm gas}}\!=\!0$ and the dust component is stationary
$\pabl{}{t}(\rho L_j)\!=\!0$, \ie the l.h.s.of the moment equations
vanish. Introducing a convective mixing on time scale $\tau_{\rm
mix}$\footnote{ Our mixing approach is very much simplified,
assuming that the gas/dust mix at height $z$ is exchanged by dust-free
gas from the deep interior of the object with element abundances
$\epsilon_j^0$ on a mixing timescale $\tau_{\rm mix}(z)$ which is
adapted to the results of 3D hydrodynamical models (Ludwig et al.
2003, 2006). Other works consider a diffusive mixing here (Ackerman \&
Marley (2001) in Cushing et al. (2007); Rossow (1978) in Warren et
al. (2007)). Interestingly, our model approach and the approach by
Ackerman \& Marley (2001) yield very comparably results f.i. regarding
the location of the cloud layer and the maximum dust-to-gas-ratio as
can be seen from a comparative study of dust cloud models (Helling et
al. 2008b).}, we have derived the following equations in Paper~III for
this case (see Eq.\,7 in Paper~III)
\begin{equation}
\label{eq:statLj}
  -\frac{d}{dz} \left(\frac{\rho_{\rm d}}{c_{\rm T}}L_{j+1}\right) 
  = \frac{1}{\xi_{\rm lKn}} \left( -\frac{\rho L_j}{\tau_{\rm mix}} 
  + \Vl^{j/3}J_\star 
  + \frac{j}{3}\,\chi^{\rm net}\,\rho L_{\rm j-1} \right).
\end{equation}
These are {\it the moment equations with respect to the total dust volume}
for dirty grains and we use them for $j\!=\!0,1,2$. As outlined in
(Eq.\,(9) in Paper~III), $\tmix$ is the timescale for mixing due to
convective motions and overshoot which decreases rapidly above the
convective layers with increasing height in the atmosphere.
However, instead of using Eq.\,(\ref{eq:statLj}) with $j\!=\!3$ for one
pure condensate, we have
to use a set of equations for dirty grains, i.e. {\it the third
dust moment equations for all volume contributions},
that is one equation for each condensate $s$ taken into
account. From (Eq.\,\ref{eq:Vsequ}) we find
\begin{equation}
\label{eq:statL4}
  -\frac{d}{dz} \left(\frac{\rho_{\rm d}}{c_{\rm T}}L_4^{\rm s}\right) 
  = \frac{1}{\xi_{\rm lKn}} \left( 
    - \frac{\rho L_3^{\rm s}}{\tau_{\rm mix}} 
    + \Vl^{\rm s} J_\star
    + \chi_{\rm net}^{\rm\,s}\,\rho L_2\right).
\end{equation}
Equations (\ref{eq:statLj}) for $j\!\in\!\{0,1,2\}$ and
Eqs.~(\ref{eq:statL4}) for $\rm s\!\in\!\{1,2,\,...\,,S\}$ (S is the
number of solid condensates taken into account) form a system of (S+3)
ordinary differential equations for the unknowns
$\{L_1,L_2,L_3,L_4^{\rm s}\}$.

The element conservation equations are not affected by the drift
motion of the dust grains. Therefore, Eq.~(29) in Paper~V remains
valid, from which we derived in the static stationary case (compare
Eq.~(8) in Paper~III)
\begin{eqnarray}
  \frac{\nH (\epsilon_i^0-\epsilon_i)}{\tmix} &=&
      \nu_{i,0}\,\Nl\,J_\star \nonumber\\[-3mm] 
  &+& \sqrt[3]{36\pi}\,\rho L_2 \sum\limits_{r=1}^R 
      \frac{\nu_{i,s} n_r^{\rm key} v^{\rm rel}_r \alpha_r}
           {\nu_r^{\rm key}} 
      \left(1-\frac{1}{\Sr\,b^{\rm\,s}_{\rm surf}}\right) \ ,
  \label{eq:verbrauch}
\end{eqnarray}
where $i$ enumerates the elements. $\Nl$ is the number of monomers in
the seed particles when they enter the size integration domain and
$\nu_{i,0}$ is the stoichiometric coefficient of the seeds (TiO$_2$
seeds: 1 for $i\!=\!\rm Ti$ and 2 for $i\!=\!\rm O$). $\nu_{i,s}$ is the
stoichiometric coefficient of element $i$ in solid material s.

The element conservation equations (Eq.~\ref{eq:verbrauch}) provide
algebraic auxiliary conditions for the ODE system
(Eqs.~\ref{eq:statLj} \& \ref{eq:statL4}) in the static stationary
case, \ie one first has to solve the system of non-linear algebraic
Eqs.\,(\ref{eq:verbrauch}) for $\epsilon_i$ at given $\{L_2,L_4^{\rm
s}\}$ (the dust volume composition $b^{\rm\,s}_{\rm surf}$ is known
from $L_4^{\rm s}$) before the r.h.s. of the ODE-equations can be
calculated. Since $J_\star$, $n_r^{\rm key}$ and in particular $\Sr$,
however, depend strongly on $\epsilon_i$, this requires a complicated
iterative procedure which creates the most problems in practise.

\subsection{Grain size distribution function}
\label{sec:gsdf}

The dust opacity calculations with effective medium theory and Mie
theory (see Sect.\,\ref{sec:srt}) require the dust particle size
distribution function $f(a)\,\rm[cm^{-4}]$ at every depth in the
atmosphere, where $a\,\rm[cm]$ is the particle radius. This function
is not a direct result of the dust moment method applied in this
paper. Only the total dust particle number density $n_d\!=\!\rho
L_0$ and the mean particle size $\langle
a\rangle\!=\!\sqrt[3]{3/(4\pi)}\,L_1/L_0$ are direct results that have
been used in Helling et al. (2006). In this paper, we reconstruct $f(a)$ 
from the calculated dust moments $L_j\,(j\!=\!1\,...\,4)$ in an approximate
way. We want to avoid the zeroth moment,
because it is only determined by a closure condition. The idea is to
introduce a suitable functional  formula for $f(a)$ with a set of
four free coefficients, and then determine these coefficients
from the known dust moments.  For further details, see
Appendix~\ref{app:sizedist}. Two possible functions for $f(a)$ are
discussed in Appendix~\ref{app:gsdf} and \ref{app:potexp}.

\subsection{Closure condition}

 Since $L_0$ appears only on the r.h.s. of Eq.~(\ref{eq:statLj})
for $j\!=\!0$, we need a closure condition in the form
$L_0\!=\!L_0(L_1,L_2,L_3,L_4)$ to solve our ODE-system. In this paper,
we use the results of the size distribution reconstruction technique
explained in Appendix~\ref{app:sizedist} in application to the double
delta-peaked size distribution function and write the zeroth dust
moment as
\begin{equation}
  \rho L_0 = N_1 + N_2 \ .
\end{equation}
where the two dust particle densities $N_1$ and $N_2\,\rm[cm^{-3}]$
are introduced in Appendix~\ref{app:gsdf}.

\begin{table*}
\caption{Chemical surface reactions $r$ assumed to form the solid materials 
   s. The efficiency of the reaction is limited by
   the collision rate of the key species, which has the lowest
   abundance among the reactants. The notation $\half$ in the
   r.h.s.~column means that only every second collision (and sticking)
   event initiates one reaction (see $\nu^{\rm key}_r$ in
   Eqs.~\ref{eq:chinet} and \ref{eq:verbrauch}). Data sources for the
   supersaturation ratios (and saturation vapour pressures): (1)
   Helling \& Woitke (2006); (2) Nuth \& Ferguson (2006); (3) Sharp \&
   Huebner (1990)}.
\label{tab:chemreak}
\centering
\resizebox{15.2cm}{!}{
\begin{tabular}{c|c|l|l}
{\bf Index $r$} & {\bf Solid s} & {\bf Surface reaction} & {\bf
Key species} \\
\hline 
1 & TiO$_2$[s]          & TiO$_2$ 
       $\longrightarrow$ TiO$_2$[s]                  & TiO$_2$ \\ 
2 & rutile              & Ti + 2 H$_2$O 
       $\longrightarrow$ TiO$_2$[s] + 2 H$_2$        & Ti     \\
3 & (1)                 & TiO + H$_2$O  
       $\longrightarrow$ TiO$_2$[s] + H$_2$          & TiO     \\ 
4 &                     & TiS + 2 H$_2$O
       $\longrightarrow$ TiO$_2$[s] + H$_2$S + H$_2$ & TiS     \\
\hline 
5 & SiO$_2$[s]          & SiO$_2$ 
       $\longrightarrow$ SiO$_2$[s]                  & SiO$_2$ \\ 
6 & silica              & SiO + H$_2$O 
       $\longrightarrow$ SiO$_2$[s] + H$_2$          & SiO     \\ 
7 &  (3)                & SiS + 2 H$_2$O 
       $\longrightarrow$ SiO$_2$[s] + H$_2$S + H$_2$ & SiS     \\
\hline 
8 & SiO[s]              & SiO 
       $\longrightarrow$ SiO[s]                  & SiO \\
9 & silicon mono-oxide  & SiO$_2$ + H$_2$ 
       $\longrightarrow$ SiO[s] + H$_2$O         & SiO$_2$   \\
10 & (2)                & SiS + H$_2$O 
       $\longrightarrow$ SiO[s] + H$_2$S         & SiS     \\
\hline   
11 & Fe[s]              & Fe 
       $\longrightarrow$ Fe[s]                  & Fe      \\ 
12 & solid iron         & FeO + H$_2$ 
       $\longrightarrow$ Fe[s] + H$_2$O         & FeO     \\
13 & (1)                & FeS + H$_2$ 
       $\longrightarrow$ Fe[s] + H$_2$S         & FeS     \\ 
14 &                    & Fe(OH)$_2$ + H$_2$ 
       $\longrightarrow$ Fe[s] + 2 H$_2$O       & Fe(OH)$_2$ \\ 
\hline 
15 & FeO[s]             & FeO 
       $\longrightarrow$ FeO[s]                  & FeO\\
16 & iron\,(II) oxide   & Fe + H$_2$O
       $\longrightarrow$ FeO[s] + H$_2$          & Fe\\
17 & (3)                & FeS + H$_2$O 
       $\longrightarrow$ FeO[s] + H$_2$S         & FeS\\
18 &                    & Fe(OH)$_2$
       $\longrightarrow$ FeO[s] + H$_2$          & Fe(OH)$_2$\\
\hline
19 & FeS[s]             & FeS
       $\longrightarrow$ FeS[s]                       & FeS\\
20 & iron sulphide      & Fe + H$_2$S
       $\longrightarrow$ FeS[s]     + H$_2$           & Fe\\
21 & (3)                & FeO + H$_2$S 
       $\longrightarrow$ FeS[s] + H$_2$O     & $\min\{$FeO, H$_2$S$\}$\\
22 &                    & Fe(OH)$_2$ + H$_2$S     
       $\longrightarrow$ FeS[s] + 2 H$_2$O   & $\min\{$Fe(OH)$_2$, H$_2$S$\}$\\
\hline
23 & Fe$_2$O$_3$[s]     & 2 Fe + 3 H$_2$O 
       $\longrightarrow$ Fe$_2$O$_3$[s] + 3 H$_2$        & $\half$Fe\\
24 & iron\,(III) oxide  & 2 FeO + H$_2$O
       $\longrightarrow$ Fe$_2$O$_3$[s] + H$_2$          & $\half$FeO\\
25 & (3)                & 2 FeS + 3 H$_2$O
       $\longrightarrow$ Fe$_2$O$_3$[s] + 2 H$_2$S + H$_2$&$\half$FeS\\
26 &                    & 2 Fe(OH)$_2$ 
       $\longrightarrow$ Fe$_2$O$_3$[s] + H$_2$O + H$_2$ & $\half$Fe(OH)$_2$\\
\hline
27 & MgO[s]            & MgO
      $\longrightarrow$ MgO[s]                        & MgO\\ 
28 & periclase         & Mg + H$_2$O 
      $\longrightarrow$ MgO[s] + H$_2$                & Mg\\
29 & (3)               & 2 MgOH
      $\longrightarrow$ 2 MgO[s] + H$_2$              & $\half$MgOH\\
30 &                   & Mg(OH)$_2$
      $\longrightarrow$ MgO[s] + H$_2$O               & Mg(OH)$_2$\\
\hline
31 & MgSiO$_3$[s]     & Mg + SiO + 2 H$_2$O 
     $\longrightarrow$ MgSiO$_3$[s] + H$_2$
                                  & $\min\{$Mg, SiO$\}$\\ 
32 & enstatite        & Mg + SiS + 3 H$_2$O 
     $\longrightarrow$ MgSiO$_3$[s] + H$_2$S + 2 H$_2$
                                  & $\min\{$Mg, SiS$\}$\\ 
33 & (3)              & 2 MgOH + 2 SiO + 2 H$_2$O
     $\longrightarrow$ 2 MgSiO$_3$[s] + 3 H$_2$    
                                  & $\min\{\half$MgOH, $\half$SiO$\}$ \\
34 &                  & 2 MgOH + 2 SiS + 2 H$_2$O
     $\longrightarrow$ 2 MgSiO$_3$[s] + 2 H$_2$S + 2 H$_2$ 
                                  & $\min\{\half$MgOH, $\half$SiS$\}$ \\
35 &                  & Mg(OH)$_2$ + SiO 
     $\longrightarrow$ 2 MgSiO$_3$[s] +  H$_2$
                                  & $\min\{$Mg(OH)$_2$, SiO$\}$ \\ 
36 &                  & Mg(OH)$_2$ + SiS + H$_2$O
     $\longrightarrow$ MgSiO$_3$[s] + H$_2$S+ H$_2$
                                  & $\min\{$Mg(OH)$_2$, SiS$\}$ \\
\hline
37 & Mg$_2$SiO$_4$[s] & 2 Mg + SiO + 3 H$_2$O
     $\longrightarrow$ Mg$_2$SiO$_4$[s] + 3 H$_2$  
                                  & $\min\{\half$Mg, SiO$\}$\\
38 & forsterite       & 2 MgOH + SiO + H$_2$O
     $\longrightarrow$ Mg$_2$SiO$_4$[s] + 2 H$_2$
                                  & $\min\{\half$MgOH, SiO$\}$\\ 
39 & (3)              & 2 Mg(OH)$_2$ + SiO 
     $\longrightarrow$ Mg$_2$SiO$_4$[s] + H$_2$O + H$_2$
                                  & $\min\{\half$Mg(OH)$_2$, SiO$\}$ \\ 
40 &                  & 2 Mg + SiS + 4 H$_2$O            
     $\longrightarrow$ Mg$_2$SiO$_4$[s] + H$_2$S + 3 H$_2$
                                  & $\min\{\half$Mg, SiS\} \\
41 &                  & 2 MgOH + SiS + 2 H$_2$O 
     $\longrightarrow$ Mg$_2$SiO$_4$[s] + H$_2$S + 2 H$_2$
                                  & $\min\{\half$MgOH, SiS\}\\
42 &                  & 2 Mg(OH)$_2$ + SiS 
     $\longrightarrow$ Mg$_2$SiO$_4$[s] + H$_2$ + H$_2$S
                                  & $\min\{\half$Mg(OH)$_2$, SiS\} \\
\hline 
43 & Al$_2$O$_3$[s]   & 2 Al + 3 H$_2$O 
     $\longrightarrow$ Al$_2$O$_3$[s] + 3 H$_2$   & $\half$Al\\
44 & aluminia         & 2 AlOH + H$_2$O 
     $\longrightarrow$ Al$_2$O$_3$[s] + 2 H$_2$   & $\half$AlOH \\ 
45 & (3)              &  2 AlH + 3 H$_2$O 
     $\longrightarrow$ Al$_2$O$_3$[s] + 4 H$_2$   & $\half$AlH\\
46 &                  & Al$_2$O + 2 H$_2$O
     $\longrightarrow$ Al$_2$O$_3$[s] + 2 H$_2$   & Al$_2$O\\ 
47 &                  & 2 AlS + 3 H$_2$O
     $\longrightarrow$ Al$_2$O$_3$[s] + 2 H$_2$S + H$_2$ & $\half$AlS\\
48 &                  & 2 AlO$_2$H 
     $\longrightarrow$ Al$_2$O$_3$[s] + H$_2$O    & $\half$AlO$_2$H\\ 
\hline
49 & CaTiO$_3$[s]         & Ca + TiO + 2 H$_2$O 
      $\longrightarrow$ CaTiO$_3$[s] + 2 H$_2$       & $\min\{$Ca, TiO$\}$\\
50 & perovskite           & Ca + TiO$_2$ + H$_2$O 
      $\longrightarrow$ CaTiO$_3$[s] + H$_2$         & $\min\{$Ca, TiO$_2\}$\\
51 &  (3)                 & Ca + Ti + 3 H$_2$O     
      $\longrightarrow$ CaTiO$_3$[s] + 3 H$_2$       & $\min\{$Ca, Ti$\}$\\  
52 &                      & CaO + Ti + 2 H$_2$O 
      $\longrightarrow$ CaTiO$_3$[s] + 2 H$_2$       & $\min\{$CaO, Ti$\}$\\
53 &                      & CaO + TiO + H$_2$O 
      $\longrightarrow$ CaTiO$_3$[s] + H$_2$         & $\min\{$CaO, TiO$\}$\\
54 &                      & CaO + TiO$_2$
      $\longrightarrow$ CaTiO$_3$[s]                 & $\min\{$CaO, TiO$_2\}$\\
55 &                      & CaS + Ti + 3 H$_2$O 
      $\longrightarrow$ CaTiO$_3$[s] + H$_2$S + H$_2$& $\min\{$CaS, Ti$\}$\\
56 &                      & CaS + TiO + 2 H$_2$O 
      $\longrightarrow$ CaTiO$_3$[s] + H$_2$S + 2 H$_2$&$\min\{$CaS, TiO$\}$\\
57 &                      & CaS + TiO$_2$ + H$_2$O 
      $\longrightarrow$ CaTiO$_3$[s] + H$_2$S        & $\min\{$CaS, TiO$_2\}$\\
58 &                      & Ca(OH)$_2$ + Ti + H$_2$O 
      $\longrightarrow$ CaTiO$_3$[s] + 2 H$_2$  & $\min\{$Ca(OH)$_2$, Ti$\}$\\
59 &                      & Ca(OH)$_2$ + TiO 
      $\longrightarrow$ CaTiO$_3$[s] + H$_2$    & $\min\{$Ca(OH)$_2$, TiO$\}$\\
60 &                      & Ca(OH)$_2$ + TiO$_2$ 
      $\longrightarrow$ CaTiO$_3$[s] + H$_2$O   &$\min\rm\{Ca(OH)_2,TiO_2\}$\\
\end{tabular}}
\end{table*}

\section{Material equations and input data} 
\label{sec:mat}

To solve our model equations, we need an atmospheric
$(T, p, {\rm v}_{\rm conv})$-structure and additional material equations to calculate the number
densities of the key reactants $n_r^{\rm key}$, the nucleation rate
$J_\star$, the reaction supersaturation ratios $\Sr$, and the sticking
probabilities $\alpha_r$.

%% solid vapour pressures $p^{\rm s}_{\rm vap}$ -> S -> \Sr
%% stoichiometric factor $\nu_r^{\rm key}$      -> trivial
 
\subsection{Atmospheric $(T,p)$-structure}

The approach of this paper is to investigate the behaviour of the
dust component in quasi-static substellar atmospheres and to
propose how our approach to treat the micro-physical processes of
the formation of composite dust grains (see Sec.~\ref{sec:dustmodel}) can be included
in the frame of stellar atmosphere codes. 

For this purpose, it is
sufficient to use a given $(T, p, {\rm v}_{\rm conv})$ structure
representing a brown dwarf or giant-gas atmosphere. The atmospheric
structure used for our models are {\it cond} AMES atmosphere
structures\footnote{ftp.ens-lyon.fr/pub/users/CRAL/fallard/}. The
feedback of the dust formation and the presence of the dust on the
atmospheric structure is thereby neglected in this paper, although the
model code to calculate the  $(T, p, {\rm v}_{\rm conv})$  structures did incorporate some dust
modelling (Allard et al. 2001). Such a decoupled approach may not be
entirely satisfying, but the understanding of the physics of cloud
layers requires some basic studies, before the feedback mechanisms
can be attacked in the frame of highly nonlinear stellar atmosphere
codes. For consistent models of our dust treatment in the frame of the
PHOENIX stellar atmosphere code including radiative transfer and
convection see ( Dehn 2007, Helling et al. 2008a). 

\subsection{Gas-phase chemistry}

We calculate the particle densities of all gaseous species, including
$n_r^{\rm key}$ as described in Paper~III according to pressure,
temperature and the calculated, depth-dependent element abundances
$\epsilon_i$ in chemical equilibrium. For the well-mixed, deep element
abundances $\epsilon^0_i$ we use solar abundances according to the
{\it cond} AMES input model.  For those elements that are not included in the
calculations (abundances are calculated for Mg, Si, Ti, O, Fe, Al, Ca,
S), we put $\epsilon_i\!=\!\epsilon^0_i$.

\subsection{Nucleation} 
\label{sec:nuc}

The nucleation rate $J_\star$ is calculated for
${\rm(TiO_2)}_N$-clusters according to Eq.~(34) in Paper V, applying
the modified classical nucleation theory of Gail et al. (1984). We use
the value of the surface tension $\sigma$ fitted to small cluster data
by Jeong~(2000) as outlined in Paper~III.

\subsection{Growth/Evaporation: 
            choice of solid material and condensation experiments}

Several dozens of solid species have been treated in phase equilibrium
in the literature (Sharp \& Huebner 1990, Fegley \& Lodders 1994). We
have taken into account 12 solids (TiO$_2$[s], Al$_2$O$_3$[s],
CaTiO$_3$[s], Fe[s], FeO[s], FeS[s], Fe$_2$O$_3$[s], SiO[s],
SiO$_2$[s], MgO[s], MgSiO$_3$[s], Mg$_2$SiO$_4$[s]) in
phase-non-equilibrium to calculate the formation and composition of
the dirty grains. Our selection is guided by the most stable
condensates which yet have simple stoichiometric ratios that ensure
that these solids can be easily built up from the gas phase. The
selection covers the main element sinks during dust formation. Our
choice is furthermore inspired by the following experiments.

\smallskip
\noindent
\underline{SiO$_2$[s], MgO[s], FeO[s], Fe$_2$O$_3$[s], MgSiO$_3$[s],
Mg$_2$SiO$_4$[s]:} Vapour phase condensation experiments offer strong
arguments against equilibrium assemblages even under controlled
conditions in the terrestrial laboratory.  Rietmeijer et al.~(1999)
have shown that from a Fe-Mg-SiO-H$_2$O vapour only $\rm
Mg_xFe_yO_z$-condensates with simple stoichiometric ratios form during
the condensation process in the laboratory, which is in accordance
with some simple phase equilibrium calculations (e.g. Sharp \& Huebner
1990), but is in contrast to well-established text books like
(Lewis~1997).  Rietmeijer et al.~(1999) report on the formation of the
end-member oxides FeO[s], Fe$_2$O$_3$[s], MgO[s] and SiO$_2$[s], and
the fosterite olivine Mg$_2$SiO$_4$[s]. 
%Hashimoto (1990) show that the
%evaporation of Mg$_2$SiO$_4$[s] favours the formation of
%SiO$_2$[s]. 
The appearance of SiO$_2$[s] shows that kinetic factors
can favour the formation of meta-stable states which is in contrast to
phase equilibrium calculations.

Ferguson \& Nuth (2006) revised the vapour pressure of SiO[s] and
discuss implications for possible $\rm(SiO)_N$ nucleation. However, the
experiments  of Rietmeijer et al. (1999) which utilise an
Fe-Mg-SiO-H$_2$O vapour do not show any formation of SiO[s]. John
(2002) pointed out that SiO[s] formation may proceed via SiO$_2$[s] +
Si[s] $\longrightarrow$ 2 SiO[s]. Since we are not yet able to treat
solid-solid reactions, we must omit these processes in our model.

\noindent
\underline{Fe[s], FeS[s]:} We add Fe[s] as high temperature condensate
and FeS[s] as  possible sulphur binding solid in accordance with
the experimental findings by Kern et al. (1993) and Lauretta et
al. (1996).

\noindent
\underline{TiO$_2$[s], Al$_2$O$_3$[s], CaTiO$_3$[s]:} $\rm(TiO_2)_N$
acts as nucleation species in our model (seed particle formation).
For consistency, TiO$_2$[s] must therefore also be included as
high-temperature solid compound to account correctly for the depletion
of Ti from the gas phase. CaTiO$_3$[s] is a very stable condensate and
is included to study the consumption of Ca from the gas phase.
According to our knowledge, no laboratory measurements on Ca-Ti oxides
are available so far. The experiments by Kern et al.~(1993) suggest
the formation of Al$_2$O$_3$[s] as high-temperature condensate.

\smallskip It is interesting to note that Beckertt \& Stolper~(1994)
 have found that by melting a CaO-MgO-Al$_2$O$_3$-SiO$_2$-TiO$_2$
system, complex compounds like CaAl$_4$O$_7$,
Ca$_3$Ti$_2$Al$_2$Si$_2$O$_{14}$ and Ca$_3$Al$_2$Si$_4$O$_{14}$ can
form. It seems logic to conclude that simpler compounds like MgO
etc. need to form before more complex compounds can be built. It might
furthermore be interesting to consider Earth-crust-like solids like
CaSiO$_3$ or CaCO$_3$. CaSiO$_3$ is very abundant on Earth but it
forms only under high-pressure conditions inside the Earth crust at
$10^4 - 10^6\,$bar (A.P. Rossi 2006, priv. com.). CaCO$_3$
(carbonaceous calcite), as all the other carbon bearing solids solids,
can only form in an oxygen-rich environment if the carbon is not
locked into the CO molecule.  Under the high pressure conditions
in brown dwarf atmospheres, this release takes place only below
$\approx 800$K under chemical equilibrium conditions. Toppani et
al.~(2005) have concluded that carbonates can only form in a
H$_2$O(g)-CO$_2$(g)-rich, high-density region under conditions far
away from thermal equilibrium.

\medskip
 Our selection of 12 solids is assumed to be formed by 60 chemical
surface reactions (see Table~\ref{tab:chemreak}). We assume for the
sticking coefficient $\alpha_r\!=\!1$ due to the lack of data (see
discussion in Paper~V). The Gibbs free energy data of the solid
compounds, from which the supersaturation ratios $S$ are calculated,
is mostly taken from Sharp \& Huebner (1990)  which were obtained
for crystalline materials. The data for TiO$_2$[s] is given in
Paper~III. The new vapour pressure data for SiO (Ferguson \& Nuth
2006) is used.  The reaction supersaturation ratios $\Sr$ are
calculated as outlined in Appendix B of Paper V.

\begin{figure}
 \centering
 \includegraphics[width=8.5cm]{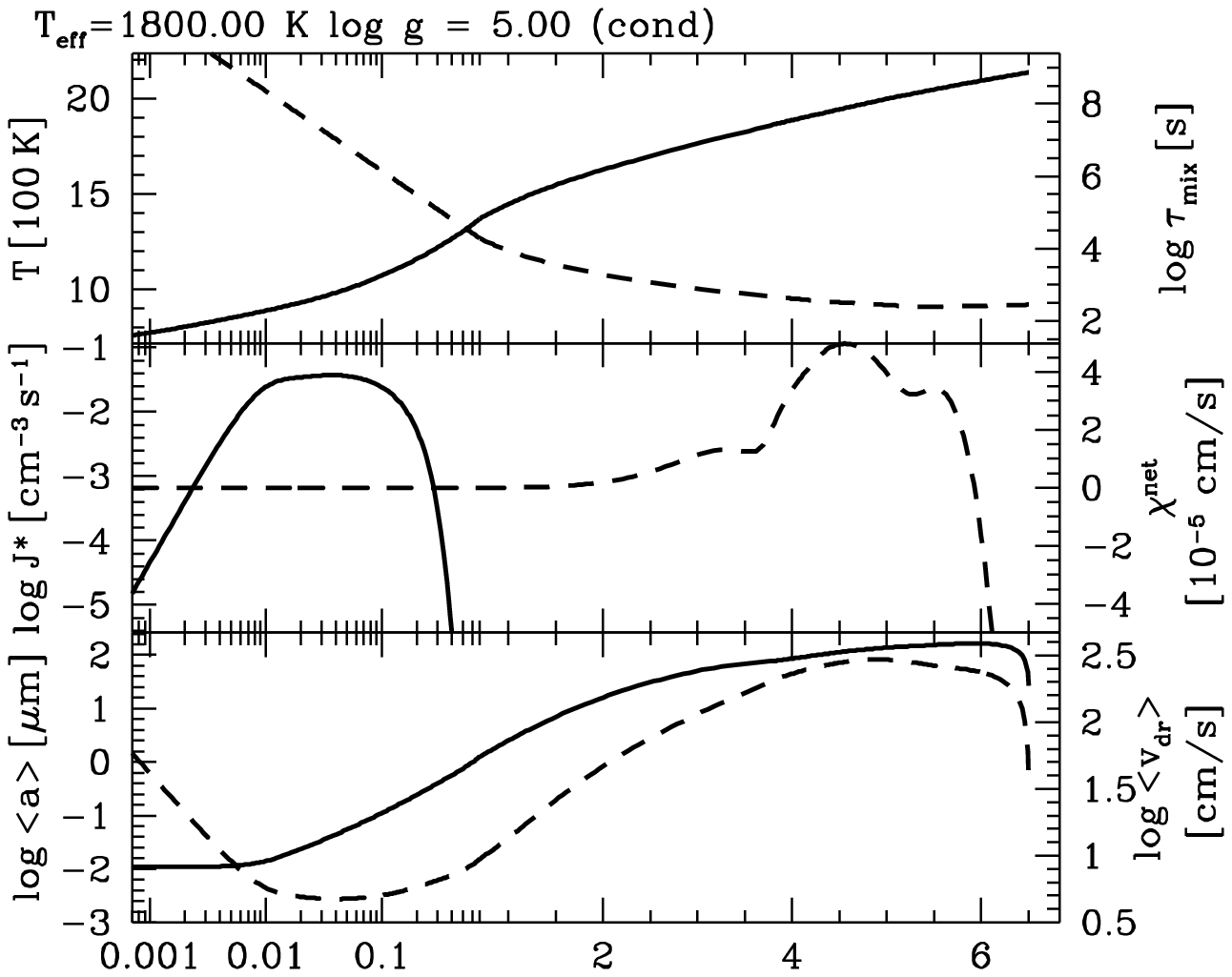}\\[1mm]
 % --- Figure 1 ---
 \hspace*{-0.3cm}\includegraphics[width=8.45cm]{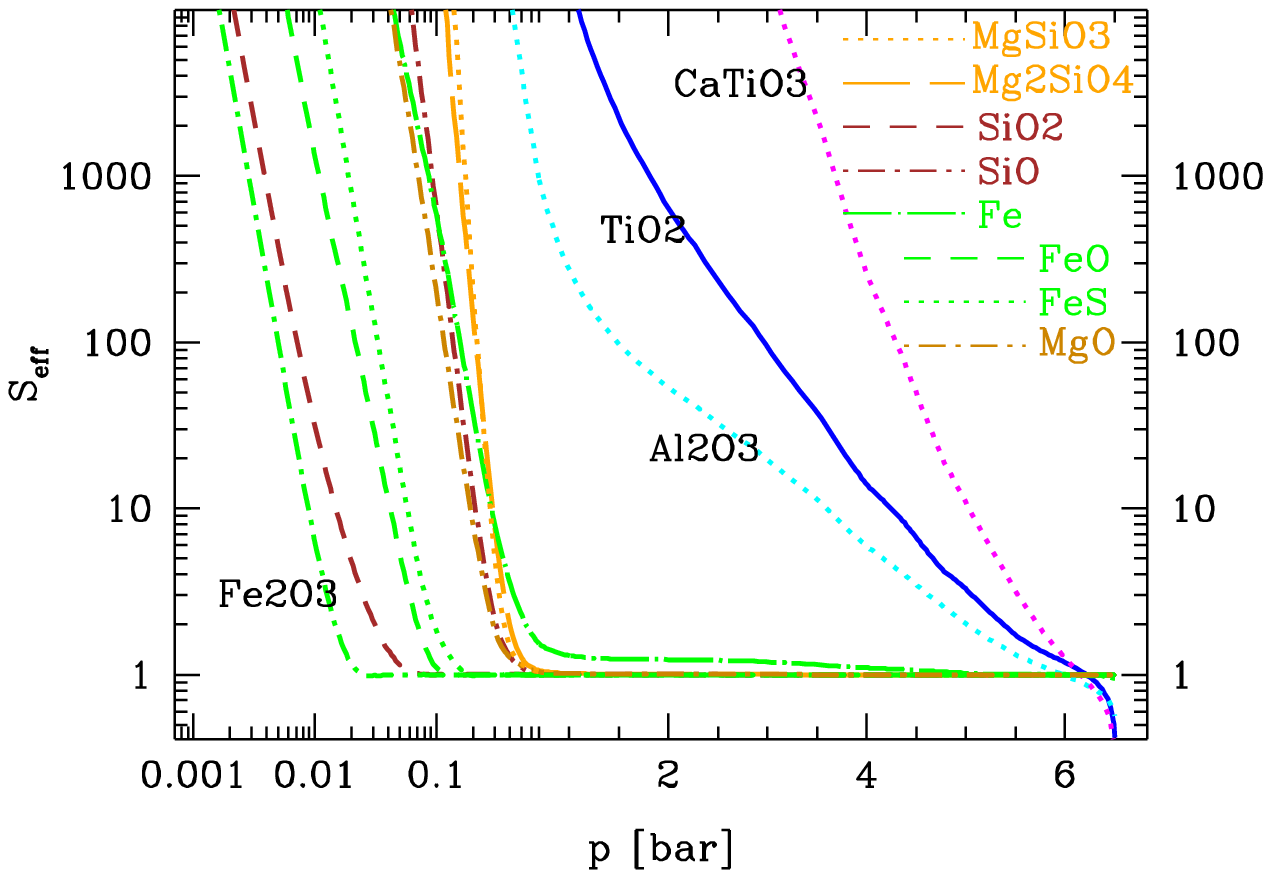}\\
   \caption{Calculated cloud structure for a model with $T_{\rm
   eff}\!=\!1800$\,K and $\log\,g\!=\!5$ ({\it cond} AMES). {\bf
   1$^{\rm st}$ panel:} prescribed temperature $T$ (solid) and mixing
   time scale $\tau_{\rm mix}$ (dashed). {\bf 2$^{\rm nd}$ panel:}
   nucleation rate $J_\star$ (solid) and net growth velocity
   $\chi_{\rm net}$ (dashed). {\bf 3$^{\rm rd}$ panel:} mean grain
   size $\langle a\rangle=\sqrt[3]{3/(4\pi)}\,L_1/L_0$ (solid) and
   mean fall speed $\langle \vdreq\rangle=\sqrt{\pi}\,g\rho_{\rm
   d} \langle a\rangle\,/\,(2\rho c_T)$ (dashed). The {\bf $4^{\rm
   th}$ panel} depicts the effective supersaturation ratios $S_{\rm
   eff}$ for all solids involved.}
 \label{Struc1800AMESCOND}
\end{figure}

\section{Results}
\label{sec:results}

\subsection{Cloud formation and cloud structure}
\label{sec:clstr}

The vertical structure of the dust cloud layer in a brown dwarf
atmosphere is depicted in Fig.~\ref{Struc1800AMESCOND}. The results
are shown for a stellar parameter combination typically associated
with the L dwarf regime.

The  cloud structure results from a hierarchical
dominance of nucleation (uppermost layers), growth \& drift
(intermediate layers), and evaporation (deepest layers). In the
uppermost layers, small grains of size $\sim\!0.01\,\mu$m form by
nucleation and grow further as they settle down the atmosphere. The
particles reach a maximum size of about $200\,\mu$m at cloud base,
shortly before they evaporate completely. The grain fall speeds first
decrease with increasing depth because of the increasing ambient gas
densities, and then re-increase as the particles grow rapidly.
Finally, they fall faster than they can grow ({\it rain}) and 
reach an approximately constant fall speed of a few m/s.  Eventually,
the grains enter the hotter atmospheric layers where they are no longer
thermally stable.  Hence, the grains shrink in size and dissolve into
the surrounding hot and convective gas  (the Schwarzschild
pressure, where the atmosphere becomes convectively unstable, is
3.54\,bar in this model). These results reflect the stationary
character of the dust component in substellar atmospheres, where
dusty material constantly forms at high altitudes and settles
downward, and simultaneously, fresh uncondensed material is mixed up by
convective motions and overshoot (see Ludwig et al. 2002, 2006; Young
et al. 2003).  These general results resemble well the results of
Paper~III, where only one sample dust species TiO$_2$[s] was
considered.

However, in comparison to Paper~III, there are new features
resulting from the inclusion of more than one solid growth species and the
consideration of the more abundant condensates.  The net growth speed
of the grains $\chi_{\rm net}$ is never determined by a single species
alone, but results from a complicated superposition where the
individual contributions $\chi_{\rm net}^{\rm s}$ can be positive
(growth) or negative (evaporation). The little kinks in $\chi_{\rm
net}$ (see 2$^{\rm nd}$ panel in Fig.~\ref{Struc1800AMESCOND}) result
from the evaporation of one material which becomes thermally unstable
at a certain temperature, in this case Mg$_2$SiO$_4$[s] at around
1800\,K and Fe[s] at around 2000\,K.

\begin{figure}
   \centering
   \includegraphics[width=8.5cm]{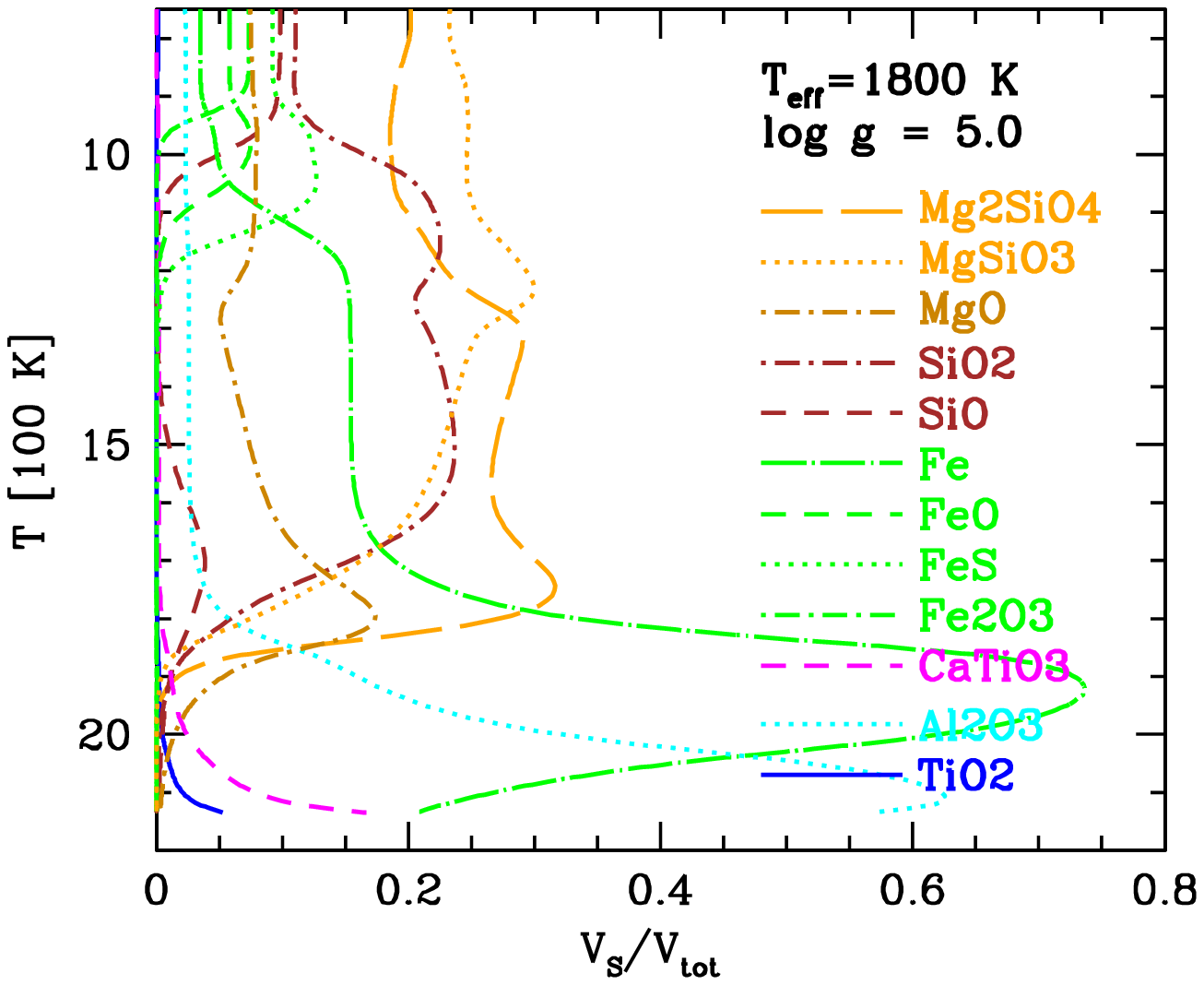}\\
   \includegraphics[width=8.5cm]{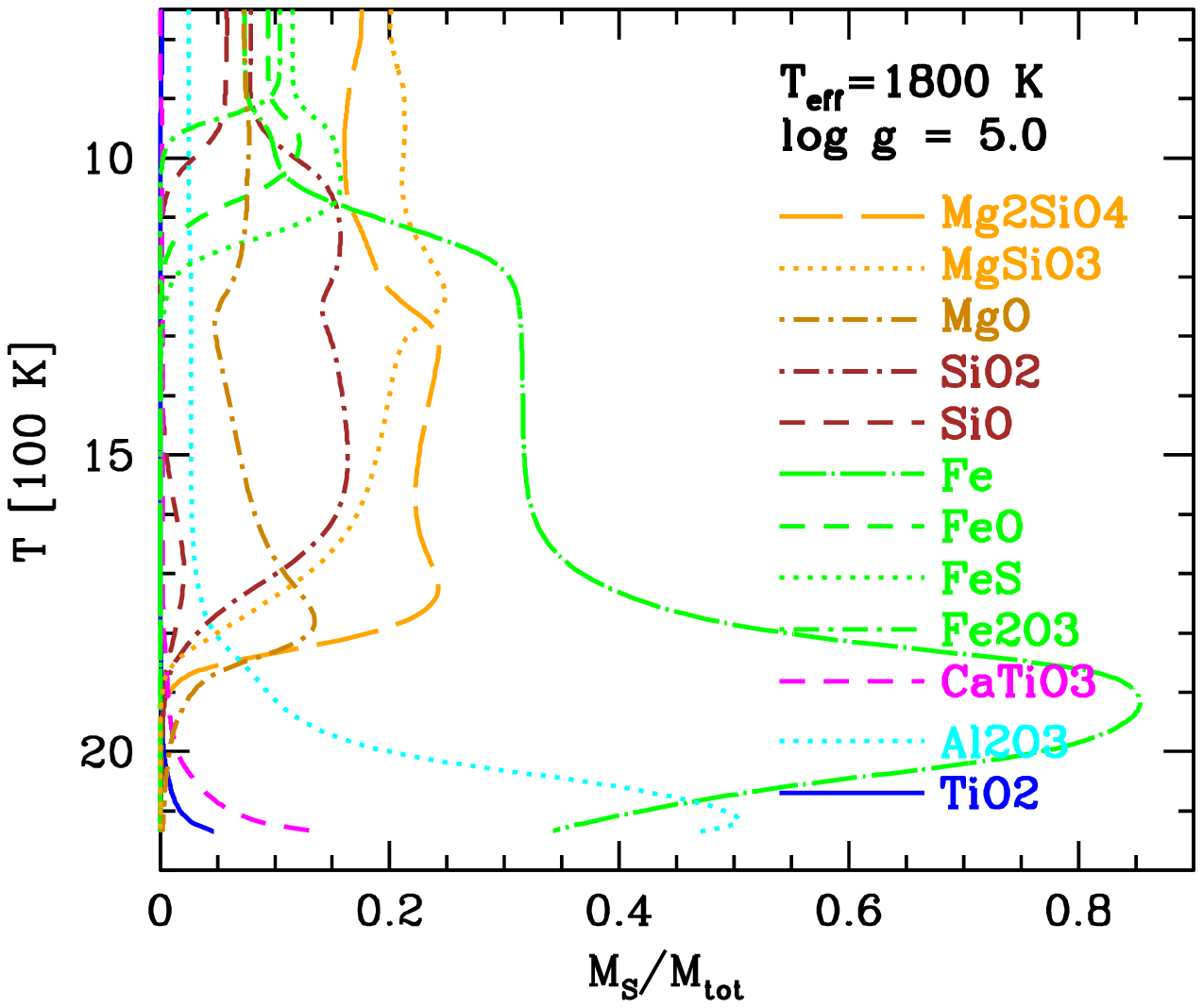}
   % --- Figure 2 ---
   \caption{Volume fractions $V_s/V_{\rm tot}$ (top) and mass
   fractions $M_s/M_{\rm tot}$ (bottom) of the various solid compounds
   for the model depicted in Fig.~\ref{Struc1800AMESCOND}. The
   temperature scale follows the atmospheric stratification being cool
   at the top and hot at the bottom.}  
   \label{fig:VOLMASSFRAC}
\end{figure}
 
\begin{figure}
   \centering
   \includegraphics[width=8.5cm]{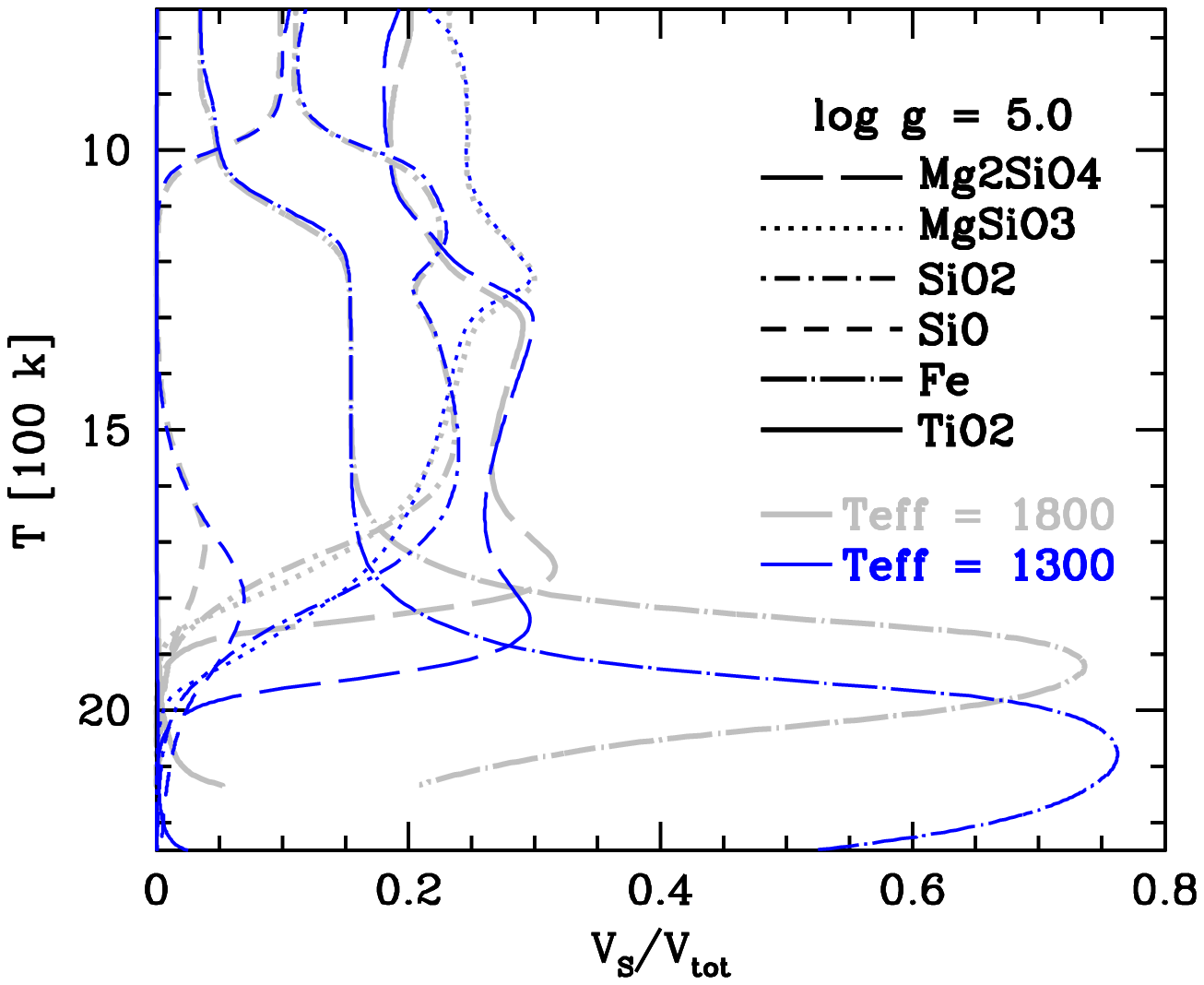}\\
   \includegraphics[width=8.5cm]{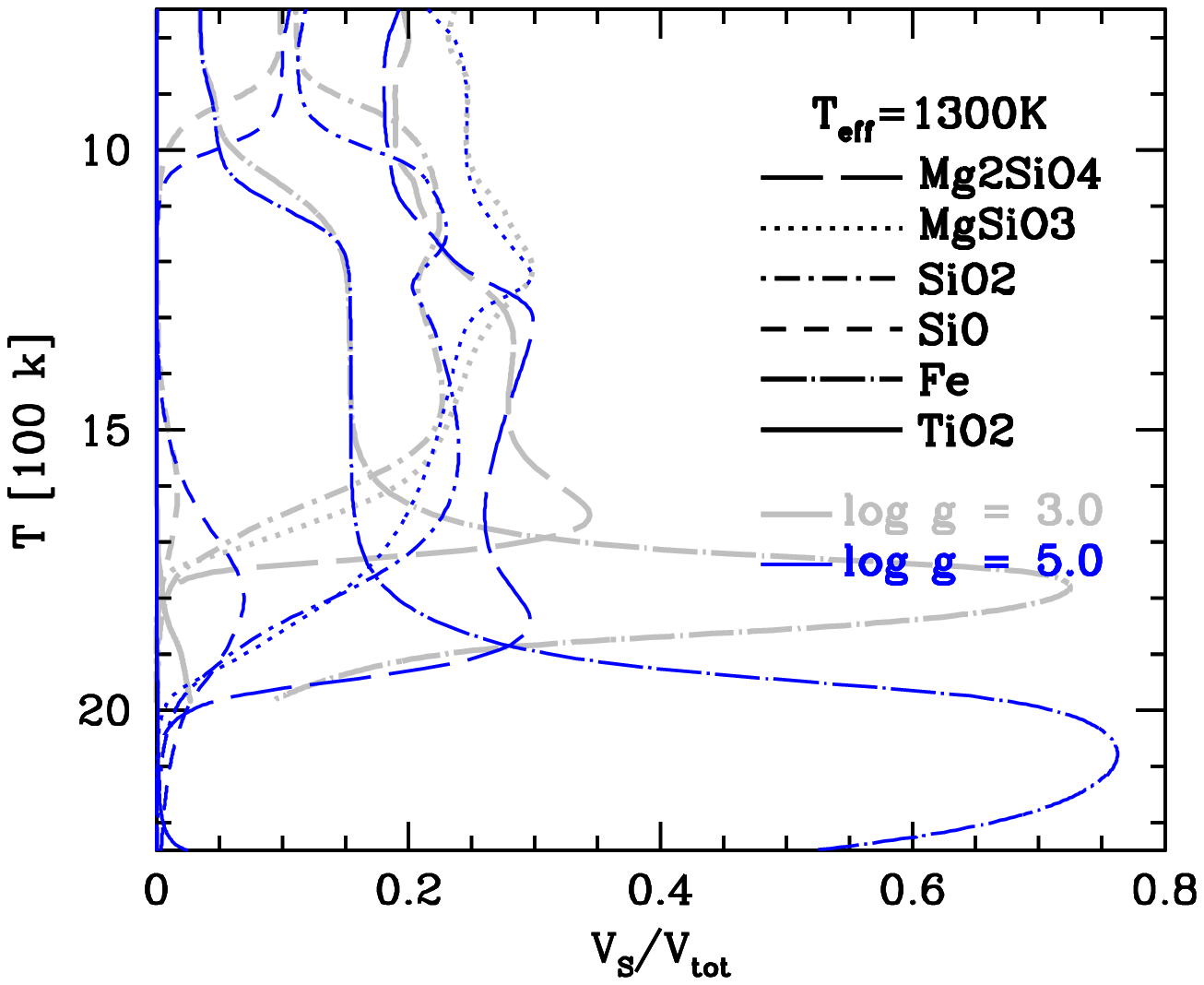}
   % --- Figure 3 ---
   \caption{Volume fractions $V_{\rm s}/V_{\rm tot}$ for different
            $T_{\rm eff}$ (top) and different $\log\,g$ (bottom). Only
            the more abundant solid compounds are viewed in contrast
            to Fig.~\ref{fig:VOLMASSFRAC}.}
   \label{fig:TeffLoggDep}
\end{figure}

\subsection{Vapour saturation in the atmosphere}

 In order to discuss deviations from phase-equilibrium, we consider
the supersaturation ratio $S$, which usually indicates net growth for
$S\!>\!1$ and net evaporation for $S\!<\!1$. However, for the
complicated growth reactions listed in Table~\ref{tab:chemreak}, where
mostly at least 2 gas particles need to collide with the grain's
surface (called ``type~III reactions'' in Paper~V), the
supersaturation ratio is not unique, but reaction-dependent (see
Appendix~B of Paper~V). In addition, the $b^{\rm\,s}_{\rm
surf}$-factors are involved to account for the inequality of the active
surfaces for growth and evaporation (see Paper~V). The thermal stability of the
solids is then better defined by $\chi_{\rm net}^{\rm\,s}\!=\!0$ (see
Eq.~\ref{eq:chinet}). In order to discuss the saturation of the
atmosphere anyway, we define a unique effective supersaturation ratio
$S_{\!\rm eff}$ for each solid as
\begin{eqnarray}
\nonumber
   \sum^R_{r=1} 
   \left(1 - \frac{1}{\Sr\,b^{\rm\,s}_{\rm surf}}\right) 
     \frac{\Delta V_r^{\rm\,s} n_r^{\rm key} {\rm v}_r^{\rm rel} 
          \alpha_r}{\nu_r^{\rm key}} =
  \\
\left(1 - \frac{1}{S^{\rm\,s}_{\rm eff}\,b^{\rm\,s}_{\rm surf}}\right) 
   \sum^R_{r=1} \frac{\Delta V_r^{\rm\,s} n_r^{\rm key} 
                     {\rm v}_r^{\rm rel} \alpha_r}
                    {\nu_r^{\rm key}}
\label{eq:effSat}
\end{eqnarray}
 These effective supersaturation ratios $S_{\!\rm eff}$ shown in
Figure~\ref{Struc1800AMESCOND} ($4^{\rm th}$ panel) demonstrate that
actually none of the considered solids obeys the condition of
phase-equilibrium ($S_{\!\rm eff}\!=\!1$) throughout the entire cloud
layer in our model. In particular, the atmospheric gas is strongly
supersaturated in the uppermost layers where the nucleation takes
place. However, in contrast to $\rm TiO_2$[s] discussed in
Paper~III, the more abundant silicates and iron compounds reach a
state of quasi-phase-equilibrium ($S_{\!\rm eff}\!\ga\!1$) already
at $p\!=\!0.05\,...\,1$bar whereas the rare condensates like
TiO$_2$[s], Al$_2$O$_3$[s], and CaTiO$_3$[s] practically never obey
the phase-equilibrium condition.  This behaviour is caused by the
different element depletion time scales being longest for the low
abundant elements as we have demonstrated in Paper V
(Sec.~4.4.3.).  Since the observable molecular features are
typically formed high in the atmosphere and even the broad dust
features in the wavelength region $7-20\,\mu$m originate from layers
$p\!\approx\!0.2\,...\,0.6\,$bar in this model (see
Sect.~\ref{sec:specs}) we conclude that the observable dust in brown
dwarf atmospheres in not in phase-equilibrium with the gas.

\begin{table}
 \caption{Typical dust volume composition $V_{\rm s}/V_{\rm tot}$,
   mass composition $M_{\rm s}/M_{\rm tot}$, and mean particles sizes
   $\langle a\rangle\,[\mu$m] as function of local temperature for models as depicted in
   Fig.~\ref{fig:VOLMASSFRAC}.
  ($\nearrow$) -- increasing in $T$-interval; 
  ($\searrow$) -- decreasing in $T$-interval.}
 \label{tab:chemcomp}
 \vspace*{-2mm}
 \begin{center}
 \begin{tabular}{r|l|l|l}
\hline
\hline
\noalign{\smallskip}
 $T$ & \multicolumn{1}{c|}{$V_s/V_{\rm tot}$} 
     & \multicolumn{1}{c|}{$M_s/V_{\rm tot}$} 
     & $\langle a\rangle$\\
 $\rm[K]$ &  &  & $\rm[\mu$m] \\
\hline
 \!700  & \!24\% MgSiO$_3$, 20\% Mg$_2$SiO$_4$\!\!   
                 & \!1. MgSiO$_3$                     & \!$10^{-3}$ \\
 \!.    & \!12\% SiO$_2$, 10\% SiO,  9\% FeS\!\!\!\! 
                 & \!2. Mg$_2$SiO$_4$                 & \!.\\
 \!.    & \!\{MgO, FeO, Fe$_2$O$_3$\} $<$ 9\%        
                 & \!3. FeS, Fe$_2$O$_3$, FeO\!\!     & \!.\\
 \!.    & \!\{Fe,  Al$_2$O$_3$\}$<$ 5\%            
                 & \!4. SiO$_2$, MgO, Fe              & \!.\\
 \!950  & \!\{TiO$_2$, CaTiO$_3$\} $<$ 1\%       
                 & \!5. SiO                           &\!$10^{-2}$\\
\hline 
 \!950  & \multicolumn{2}{c|}{\ }                     &\!$10^{-2}$\\
 \!.    & \multicolumn{2}{c|}{strongly changing}      & .\\
 \!1200 &  \multicolumn{2}{c|}{\ }                    &\!$10^{-0.5}\!\!$\\
\hline
 \!1200 & \!35\% Mg$_2$SiO$_4$, 23\% SiO$_2$        
                 & \!1. Fe                            &\!$10^{-0.5}\!\!$\\
 \!.    & \!$<$ 20\% MgSiO$_3$ ($\searrow$)         
                 & \!2. Mg$_2$SiO$_4$                 & \!.\\
 \!.    & \!15\% Fe, 5\% MgO                        
                 & \!3. SiO$_2$/MgSiO$_3$             & \!.\\
 \!1700 & \!everything else $<$5\%                  
                 & \!4. MgO                           & \!$10$\\
\hline
 \!1700 & \multicolumn{2}{c|}{\ }                   & \!$10$ \\
 \!.    & \multicolumn{2}{c|}{strongly changing}    & \!. \\
 \!1900 & \multicolumn{2}{c|}{(SiO, MgO peaking but low \%)} 
                                                    &\!$10^2$ \\
\hline
 \!1900 & \!72\% Fe ($\searrow$)        
             & \!1. Fe                              & \!$10^2$ \\
 \!.    & \! 20\% Al$_2$O$_3$ ($\nearrow$), $<$5\% TiO$_2$ 
             & \!2. Al$_2$O$_3$ ($\nearrow$)        & $(\star)$ \\
 \!2100 & \! 10\% CaTiO$_3$ (at 2100K)       
             & \!3. CaTiO$_3$                       & \!0\\
\hline
\end{tabular}
\end{center}
{\ }\\*[-2ex]
  {\small $(\star)$ Mean grain sizes $\langle a\rangle$ can reach up to
  $10^{3.5}$ in lower gravity atmospheres like giant-gas planets 
  (compare Fig.~\ref{J*aquer}).}
\vspace*{-1mm}
\end{table}

\subsection{Chemical composition of the cloud particles}

The chemical composition of the dust grains is expressed  by the
solid material volume fractions $V_{\rm s}/V_{\rm tot}$
(Fig.~\ref{fig:VOLMASSFRAC}, top panel) and mass fractions $M_{\rm
s}/M_{\rm tot}\!=\!V_{\rm s}\,\rho_{\rm s}/(V_{\rm tot}\,\rho_{\rm
d})$ (lower panel).  The dust material composition is mainly
controlled by the temperature, and shows a constantly re-occurring
pattern in models with different stellar parameter for brown dwarfs
and gas giant-planets (see also Table~\ref{tab:chemcomp}). Once a
solid becomes thermally unstable, the respective elements evaporate
into the gas, making them available again to form other, more stable
condensates. In this way, we find a complicated mix of all condensates
high in the atmosphere with volume fractions resulting from
kinetic constrains that favour the formation of the more abundant
silicates and Fe-oxides. As these dirty particles settle down the
atmosphere they stepwise purify, until only the most stable component
parts like Al$_2$O$_3$[s], Fe[s], and CaTiO$_3$[s] remain.

Therefore, two classes of solids  can be distinguished: the {\sf 
high-temperature condensates} Fe[s] and Al$_2$O$_3$[s] with some
contributions of Ca-Ti-oxides in the deeper layers, and the {\sf
medium-temperature condensates} MgSiO$_3$[s], Mg$_2$SiO$_4$[s] and
SiO$_2$[s] with some SiO[s] and FeS[s] in the upper layers.

 As the temperature increases, SiO[s] starts to evaporate and
SiO$_2$[s] becomes the second most abundant solid material by
volume fraction. Next, MgSiO$_3$[s] evaporates which sets free some
Mg and Si to form more Mg$_2$SiO$_4$[s] and SiO$_2$[s], making
SiO$_2$[s] again the second most abundant solid. The transition from
the medium-temperature to the high-temperature composition is
characterised  by the evaporation of Mg$_2$SiO$_4$[s], which
leaves only Fe[s], Al$_2$O$_3$[s], TiO$_2$[s] and CaTiO$_3$[s] as
stable condensates. As Fe[s] evaporates, the cloud base is soon reached
where eventually the remaining Al-Ca-Ti oxides evaporate.

The volume fractions are important for the dust opacities
(see Sect.~\ref{sec:srt}), and in the observable layers the
Mg-silicates turn out to be the most relevant condensates. For
completeness we note, however, that the mass fraction $M_{\rm
s}/M_{\rm tot}$ is dominated by Fe[s] from about $1200\,$K to
$2000\,$K.

We recognise that the chemical composition of the dust grains does to
some extend depend on the completeness of the selection of solids and
on the availability of material quantities (like e.g. the sticking
coefficients as shown in Sect. 4.6. in Paper V). For example, the
partial volume of Mg$_2$SiO$_4$[s] and MgSiO$_3$[s] decrease if MgO[s]
is included, and the partial volume of SiO$_2$[s] decreases if SiO[s]
is included. For this reason, the partial volume of SiO$_2$[s] is
smaller than described in Helling et al. (2006).

\begin{figure}
\hspace*{-0.8cm}
  \includegraphics[width=11cm]
    {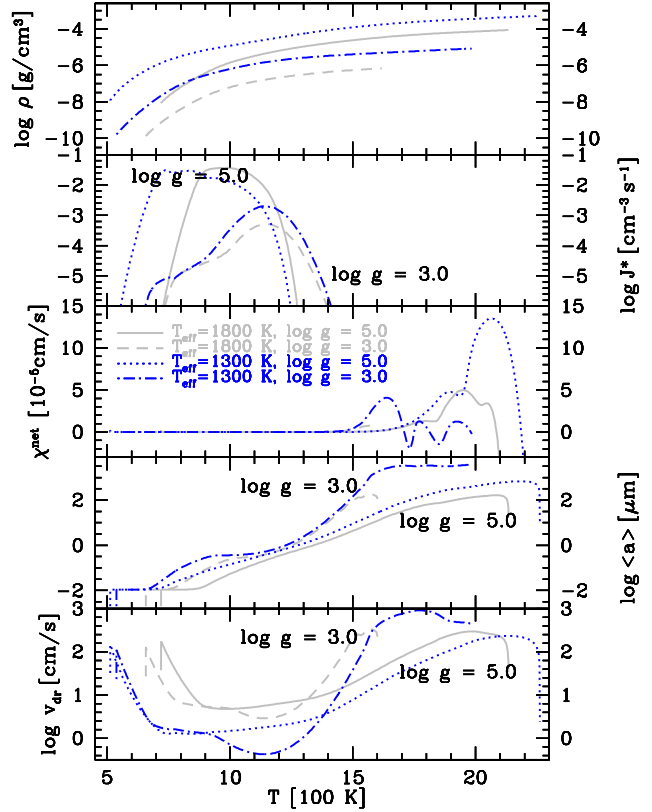}\\[0.2cm]
  % --- Figure 4 ---
  \caption{Comparison of results for different stellar parameter.
       {\bf $1^{\rm st}$ box:} mass density $\rho$,
       {\bf $2^{\rm nd}$ box:} nucleation rate $J_\star$, 
       {\bf $3^{\rm rd}$ box:} net growth velocity $\chi^{\rm net}$
%dust particle number density $n_d$, 
       {\bf $4^{\rm th}$ box:} mean particle size $\langle a\rangle$, 
       {\bf $5^{\rm th}$ box:} mean fall speed $\langle \vdreq\rangle$.
       All quantities are plotted as function of temperature $T$.
       Dotted line:      $T_{\rm eff}\!=\!1300\,$K, $\log\,g\!=\!5$;
       solid line:       $T_{\rm eff}\!=\!1800\,$K, $\log\,g\!=\!5$;
       dash-dotted line: $T_{\rm eff}\!=\!1300\,$K, $\log\,g\!=\!3$;
       dashed line:      $T_{\rm eff}\!=\!1800\,$K, $\log\,g\!=\!3$.}
  \label{J*aquer}
\end{figure}

\subsection{Dependence on $T_{\rm eff}$ and $\,\log\,g$}
\label{ssec:teff}

Figure \ref{fig:TeffLoggDep} shows the dependence of the dust
material composition on $T_{\rm eff}$ and $\log\,g$. If plotted
against temperature, the volume fractions show a robust pattern for
all calculated models (see Table~\ref{tab:chemcomp}). Only a slight
shift to higher temperatures (deeper layers) can be noticed for lower
$T_{\rm eff}$ (upper plot) and higher $\log\,g$ (lower plot).

An increase of $\log\,g$ results in a more compact atmosphere,
\ie generally higher pressures in the atmosphere. Since the
dust sublimation temperatures increase with increasing pressure, the
dust remains stable to even higher temperatures for larger
$\log\,g$. For the same reason, models with lower $T_{\rm eff}$ show
dust at comparably higher temperatures.  For lower $T_{\rm eff}$, a
certain temperature is reached deeper inside the atmosphere, where the
pressure is higher and, hence, the dust is more stable.

Figure \ref{J*aquer} shows the dependence of the nucleation rate,
the mean particle size and the mean fall speed on $T_{\rm eff}$ and
$\log\,g$. The nucleation zone lies typically between 600\,K and
1400\,K in all models. The maximum rate reaches higher values for
higher $\log\,g$ because of the higher gas densities.  For lower
$\log\,g$, the nucleation maximum is more extended and shows a more
complicated shape that is probably caused by the closer neighbourhood
to the convective zone, which makes the up-mixing of fresh elements for
nucleation more likely.

Concerning the mean particle size, all models show about the same
small particles of order $0.01\,\mu$m high in the atmosphere, which
grow to large particles between about $100\,\mu$ and $1000\,\mu$ in
the deep layers. The maximum particle sizes reached at cloud base are
larger for lower $\log\,g$. As the nucleation zone ends, there is a
zone of rapid grain growth around 1400\,K to 1700\,K. In this zone,
most of the solid particles are actually ``mixed away'' according to
our simple approach to treat the mixing by convective motions and
overshoot. The small number of particles that stay, however, settle
down deeper into a denser environment where elements are mixed up with
higher efficiency due to the decreasing distance from the convection
zone. The models show that this small number of growing particles is
sufficient to maintain a state close to phase equilibrium with the gas
concerning most elements, \ie the growth is exhaustive.  Thereby, the
particles grow further along their way down the atmosphere while their
number is ever decreasing due to lethal mixing.

 The growth of the particles, however, is limited by their own
fall speed which increases with size. The grains eventually reach a
size where their residence time is so small that further growth
becomes negligible. The residence time scale $\tau_{\rm
sink}\!=\!\vdreq/H_p$, however, depends on the atmospheric scale
height $H_p$ which is 100 times larger for the $\log\,g\!=\!3$
models. Therefore, the dust particles have more time to grow in the
low $\log\,g$ models, producing larger sizes and larger fall speeds.

 We note that the Knudsen numbers fall short of unity in the
deeper layers, for particles larger than about $1\,\mu$m in the
$\log\,g\!=\!5$ models and about $10\,\mu$m in the $\log\,g\!=\!3$
models. This means, that the frictional force and the growth velocity
should be calculated for the case of small Knudsen numbers in the
deeper layers, making necessary a Knudsen number fall differentiation
(see Paper~II for details). Therefore, the results of this paper
concerning the particle sizes in the deeper layers must be taken with
care. The true fall speeds are probably larger there and the true
growth velocities are probably smaller as compared to our results, \ie we
expect that $\langle a\rangle$ remains smaller in the deeper layers as
compared to Fig.~\ref{J*aquer}.

\begin{figure}
%   \centering
\hspace*{-0.35cm} 
  \includegraphics[width=8.8cm]{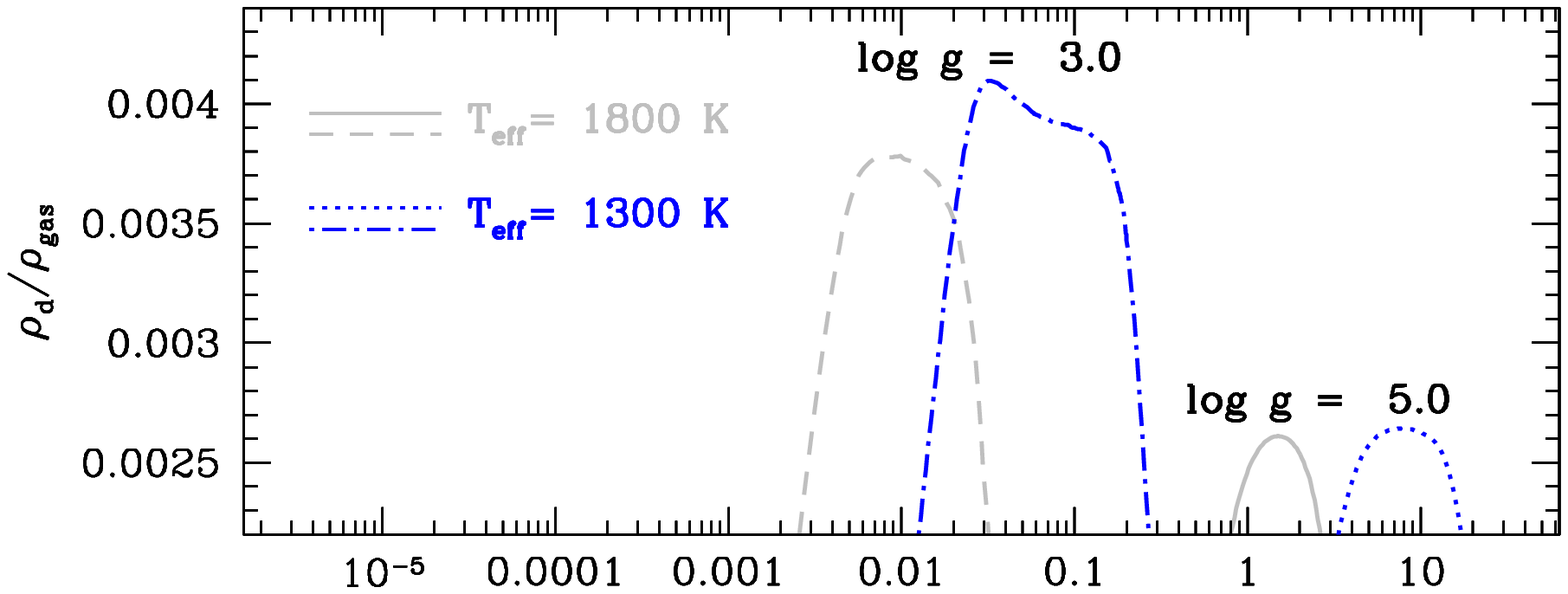}\\[0.2cm]
   \includegraphics[width=8.5cm]{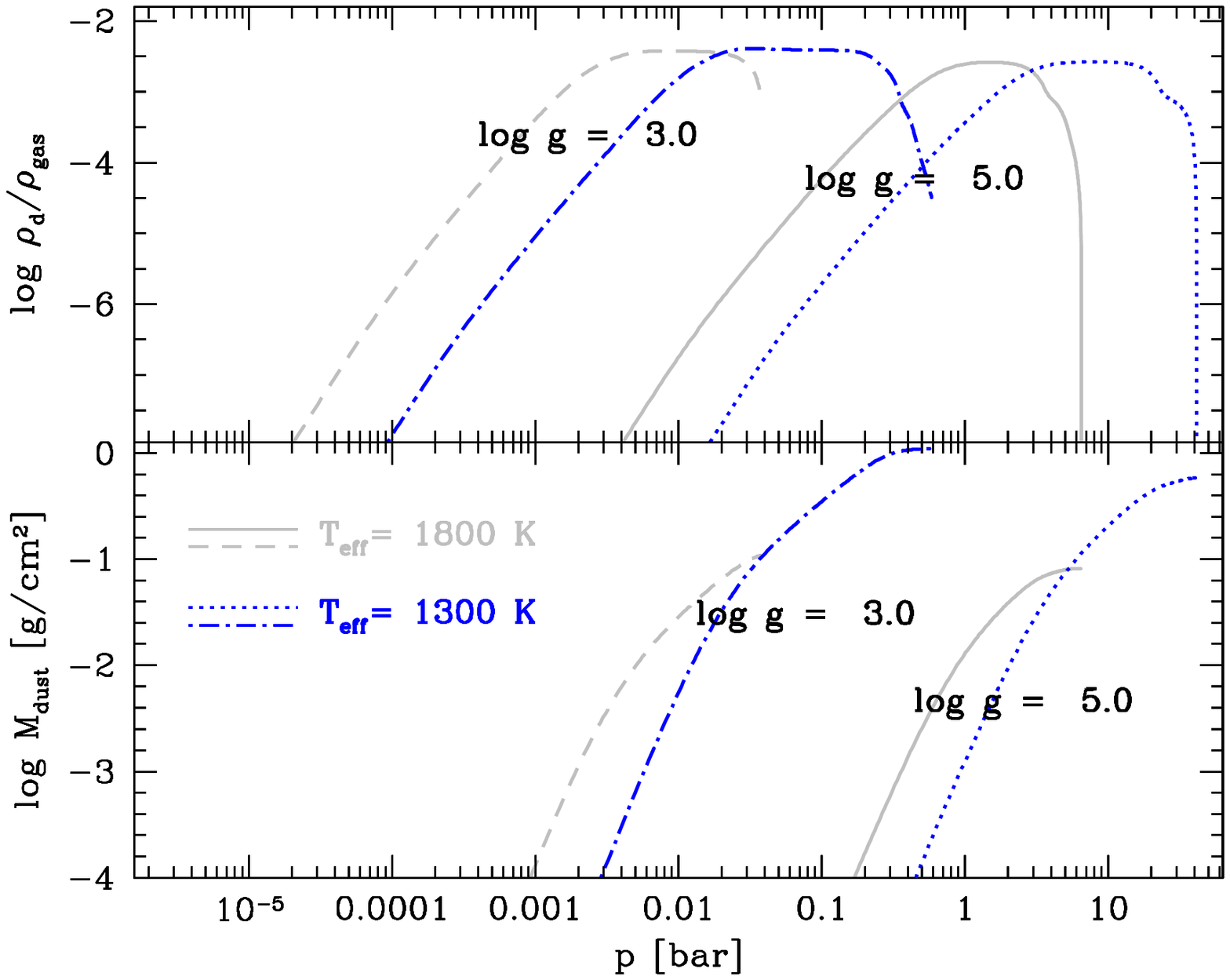}
   % --- Figure 5 ---
   \caption{Dust-to-gas mass ratios $\rho_{\rm d}/\rho_{\rm gas}$ 
      and dust mass column densities $M_{\rm dust}$ for brown
      dwarf and giant-planets atmospheres.
      {\bf $1^{\rm st}$ panel:} linear plot, 
      {\bf $2^{\rm nd}$ panel:} logarithmic plot.
      Dotted line:      $T_{\rm eff}\!=\!1300\,$K, $\log\,g\!=\!5$;
      solid line:       $T_{\rm eff}\!=\!1800\,$K, $\log\,g\!=\!5$;
      dash-dotted line: $T_{\rm eff}\!=\!1300\,$K, $\log\,g\!=\!3$;
      dashed line:      $T_{\rm eff}\!=\!1800\,$K, $\log\,g\!=\!3$.}
   \label{rhodrhog00}
\end{figure}

\begin{table}
  \centering
  \caption{Maximum dust-to-gas ratio and temperature interval where
           the dust-to-gas ratio is larger than half maximum.}
  \vspace*{-1mm}
  \label{tab:Twindow}
  \begin{tabular}{cc|cc}
  \hline
  \hline
  $T_{\rm eff}\,$[K] & $\log\,g$ & 
       $\bigg(\displaystyle\frac{\rho_{\rm d}}{\rho_{\rm gas}}\bigg)_{\rm max}$
  % \big[^o/_{oo}\big] 
  $ \big[10^{-3}\big] $ & temp.~interval\,[K] \\
  \hline
  1300 & 5 & 2.6  & $1300-1800$\\
  1800 & 5 & 2.6  & $1300-1800$\\
  2200 & 5 & 2.1  & $1450-1750$\\
  2500 & 5 & 0.2  & $1550-1800$\\
  \hline
  1300 & 3 & 4.0  &\ $950-1600$\\
  1800 & 3 & 3.8  & $1000-1600$\\
  2200 & 3 & 2.3  & $1300-1500$\\
  2500 & 3 & 0.04 & $1550-1700$\\
  \hline
  \end{tabular}
\end{table}

\subsection{Dust-to-gas ratio}

Figure~\ref{rhodrhog00} shows the dust-to-gas mass ratios $\rho_{\rm
d}/\rho_{\rm gas}$ and the dust mass column densities $M_{\rm
dust}=\int\rho_{\rm d}\,dz$ [g/cm$^2$] as function of pressure.  The
dust-to-gas ratio generally first increases inward log-linear, then
reaches a plateau and finally decreases rapidly as the dust becomes
thermally unstable. The upper plot shows the same quantities on a
linear scale, more emphasising the truly dusty layers.

 For the four models depicted in the previous figures, the
dust-to-gas ratio reaches a constant maximum value between about
$0.25\%$ and $0.4\%$ in a temperature interval
$T\!\in\!\rm[(1000-1300)\,K\,...\,(1600-1800)\,K]$ rather independent
of $T_{\rm eff}$, where the temperature interval boundaries increase
with increasing $\log\,g$ (see Sect.~\ref{ssec:teff}). The width of
the {\it dusty temperature window} is about $500\,$K to $600\,$K for the
depicted models with $T_{\rm eff}\!=1300\,$K and $1800\,$K (compare
Table~\ref{tab:Twindow}). We note that this result is actually quite
consistent with the simple dust approach made by Tsuji (2002),
although the window is broader than assumed by Tsuji according to our
findings.  Similar maximum values are found even if we increase
$T_{\rm eff}$ to 2200\,K. However, if we increase $T_{\rm eff}$
further to 2500\,K, the maximum dust-to-gas ratio drops by more than
one order of magnitude (see Table~\ref{tab:Twindow}). We conclude that
$T_{\rm eff}\!\approx\!2200\,K$ is the threshold value for truly
dust-rich layers to occur in the models presented here.

The resulting dust-to-gas ratios demonstrate that the cloud layer is
primarily attached to the local temperature. For lower $T_{\rm eff}$,
the dust layer sinks in deeper into the atmosphere and eventually
disappears from the observable layers. Since the dust is then present
in denser regions, the dust mass column density $M_{\rm dust}$
increases.

\begin{figure}
  \includegraphics[width=8.5cm]{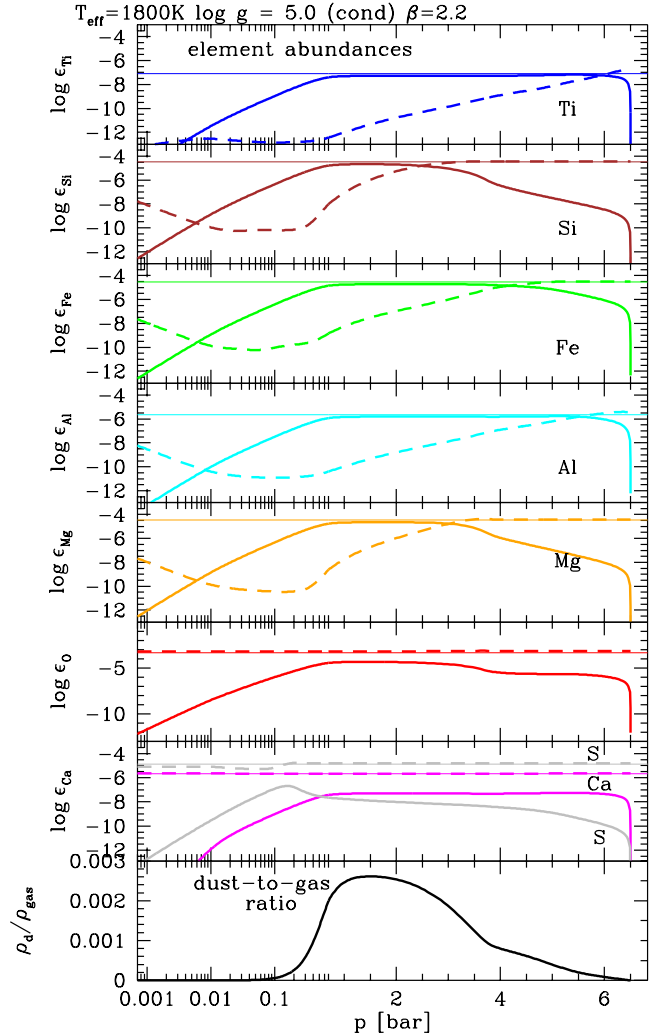}
  % --- Figure 6 ---
  \caption{Element abundances in dust $\epsilon_{\rm d}$ (thick solid)
    and in the gas phase $\epsilon_{\rm i}$ (dashed). The solar values
    $\epsilon_{\rm i, Sun}$ (thin solid) and the linear dust-to-gas
    ratio $\rho_{\rm d}/\rho_{\rm gas}$ (lowest panel) are shown for
    comparison. The stellar parameter are $T_{\rm eff}\!=\!1800\,$K
    and $\log\,g\!=\!5$ like in Fig.~\ref{Struc1800AMESCOND}}
 \label{epsi1800}
\end{figure}

\begin{figure}
  \includegraphics[width=8.5cm]{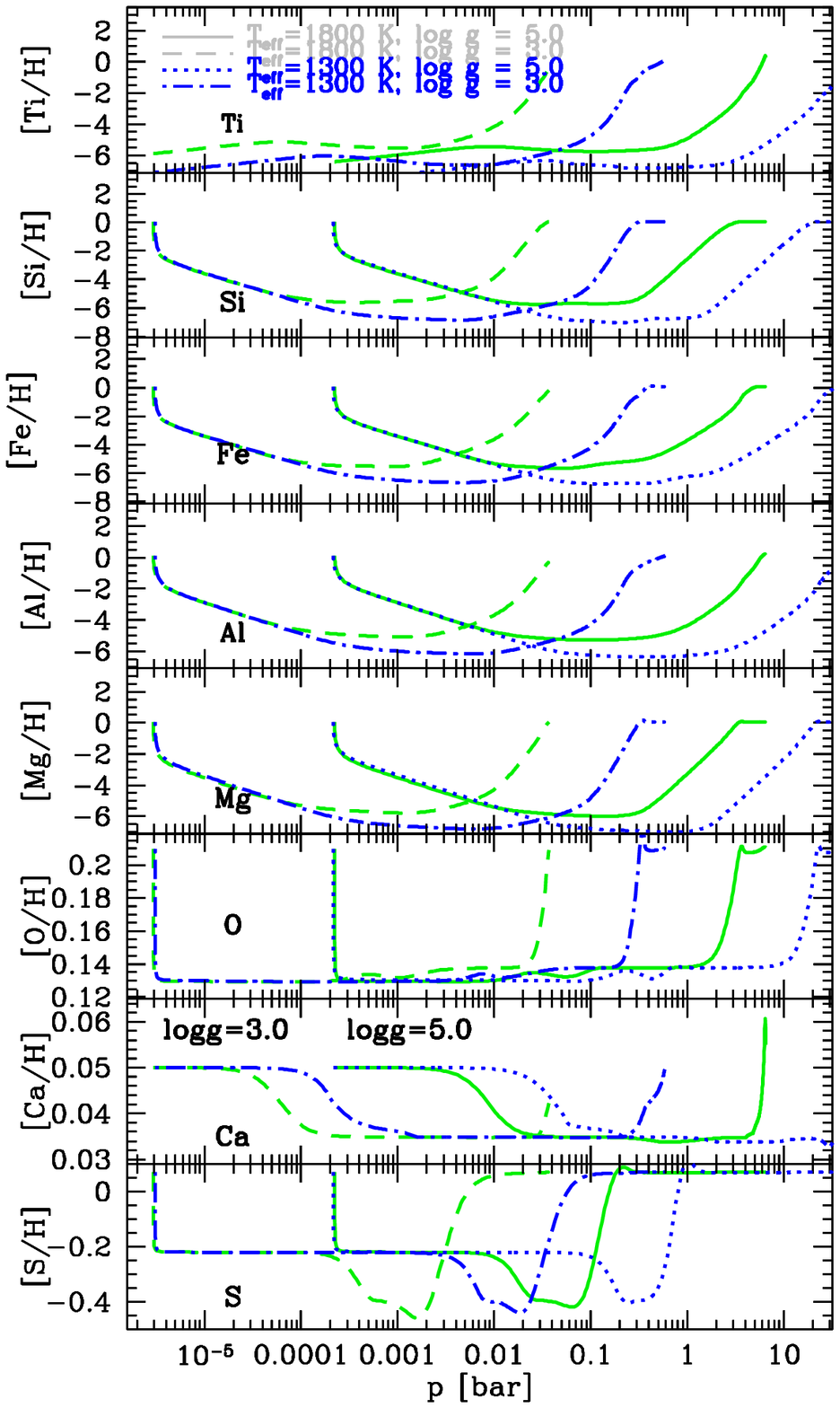}
  % --- Figure 7 ---
  \caption{Element abundances involved in the dust formation process
    for two effective temperatures and two different gravities.
    dotted blue:      $T_{\rm eff}\!=\!1300\,$K, $\log\,g\!=\!5$;
    solid green:      $T_{\rm eff}\!=\!1800\,$K, $\log\,g\!=\!5$;
    dash-dotted blue: $T_{\rm eff}\!=\!1300\,$K, $\log\,g\!=\!3$;
    dashed green:     $T_{\rm eff}\!=\!1800\,$K, $\log\,g\!=\!3$.}
 \label{metal1800}
\end{figure}

\subsection{Remaining gas-phase chemistry \& metallicity}

The remaining gas-phase composition depends on the amount of elements
not locked up into dust grains. As a result of our model, 
phase-equilibrium is not valid in the upper layers (see
Fig.~\ref{Struc1800AMESCOND}, 4$^{\rm th}$ panel)  and should not
be used to determine gas particle abundances. The remaining gas
abundances $\epsilon_{\rm i}$ are strongly sub-solar in an extended
layer above the cloud layer where the pressure drops by about 3 orders
of magnitude (about 10 scale heights), see Figs.~\ref{epsi1800} and
\ref{metal1800}. Inside the cloud layer, the metal abundances increase
with increasing atmospheric depth and finally reach  even slightly
larger than solar values at cloud base, where those elements are
released by evaporation that have been locked up into grains in the
upper layers.

Figure~\ref{epsi1800} shows this {\it phase lag} between metal gas
abundances and dust abundances clearly. Considering a path from the
top to the bottom of the atmosphere, first the metals disappear, then
the dust appears, then the metals reappear and finally the dust
disappears (the sum of dust and gas abundances is not constant).

\begin{figure*}
  \centering
  \includegraphics[width=8.5cm]{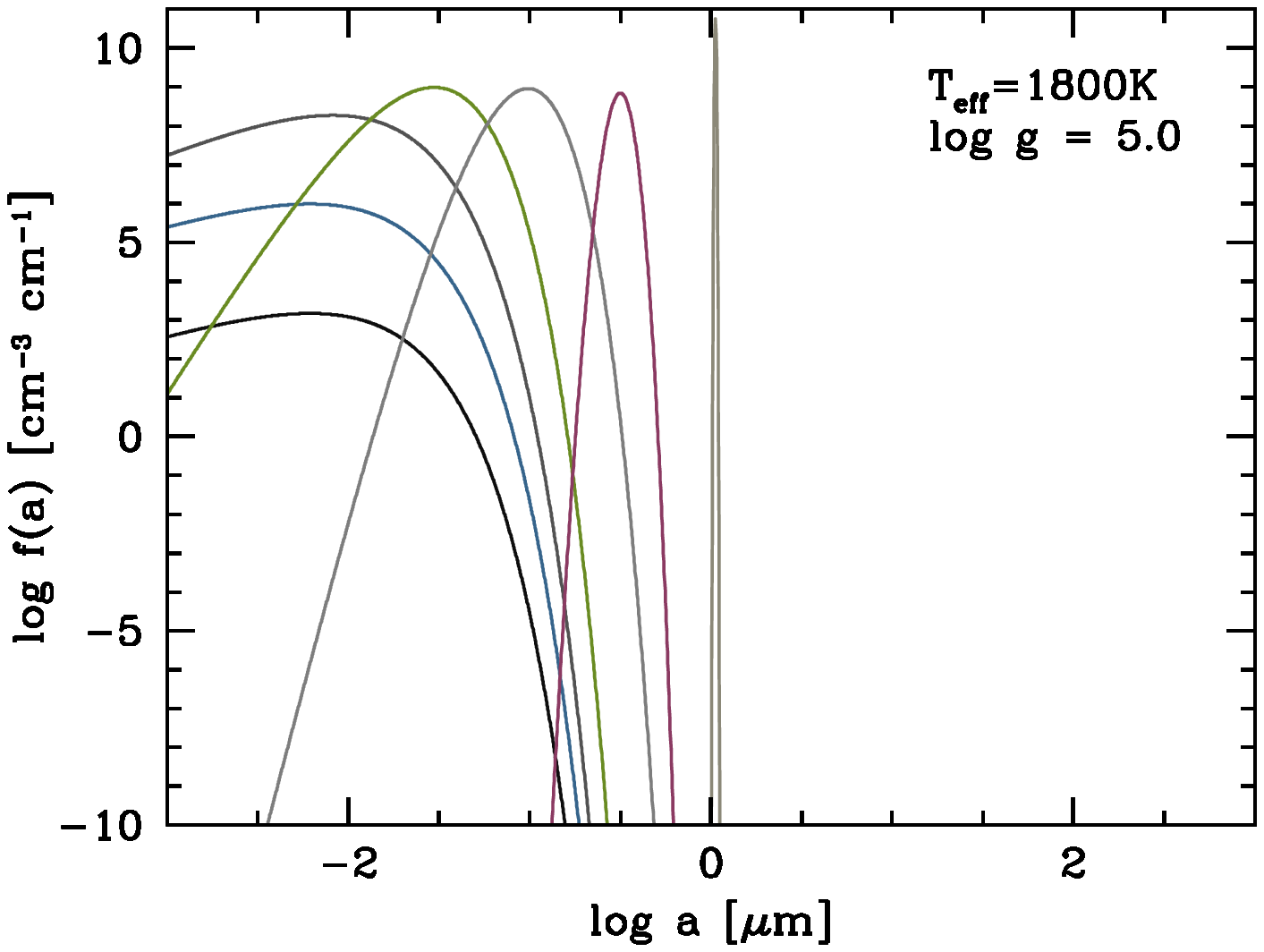}
  \includegraphics[width=8.5cm]{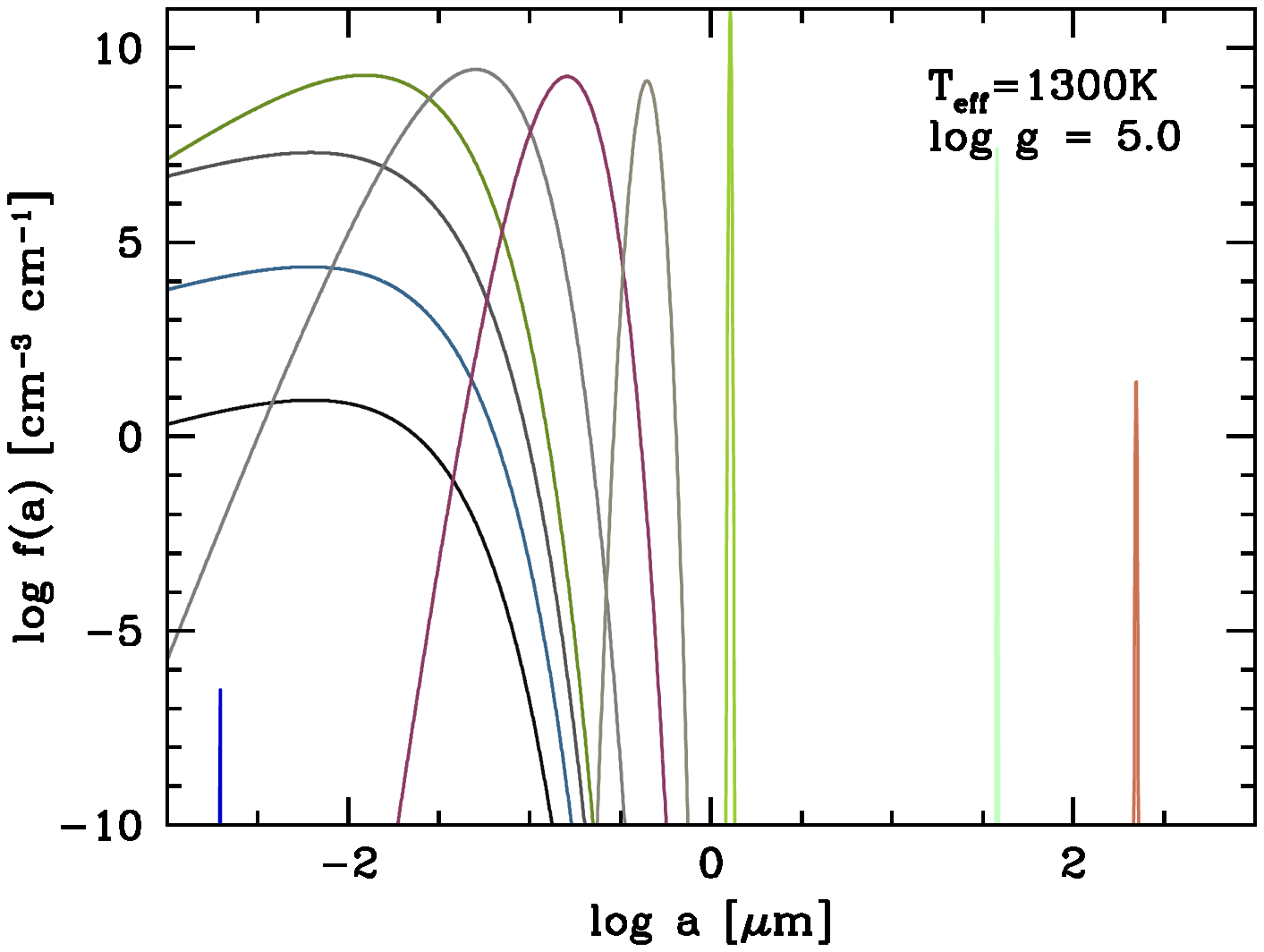}\\
  \includegraphics[width=8.5cm]{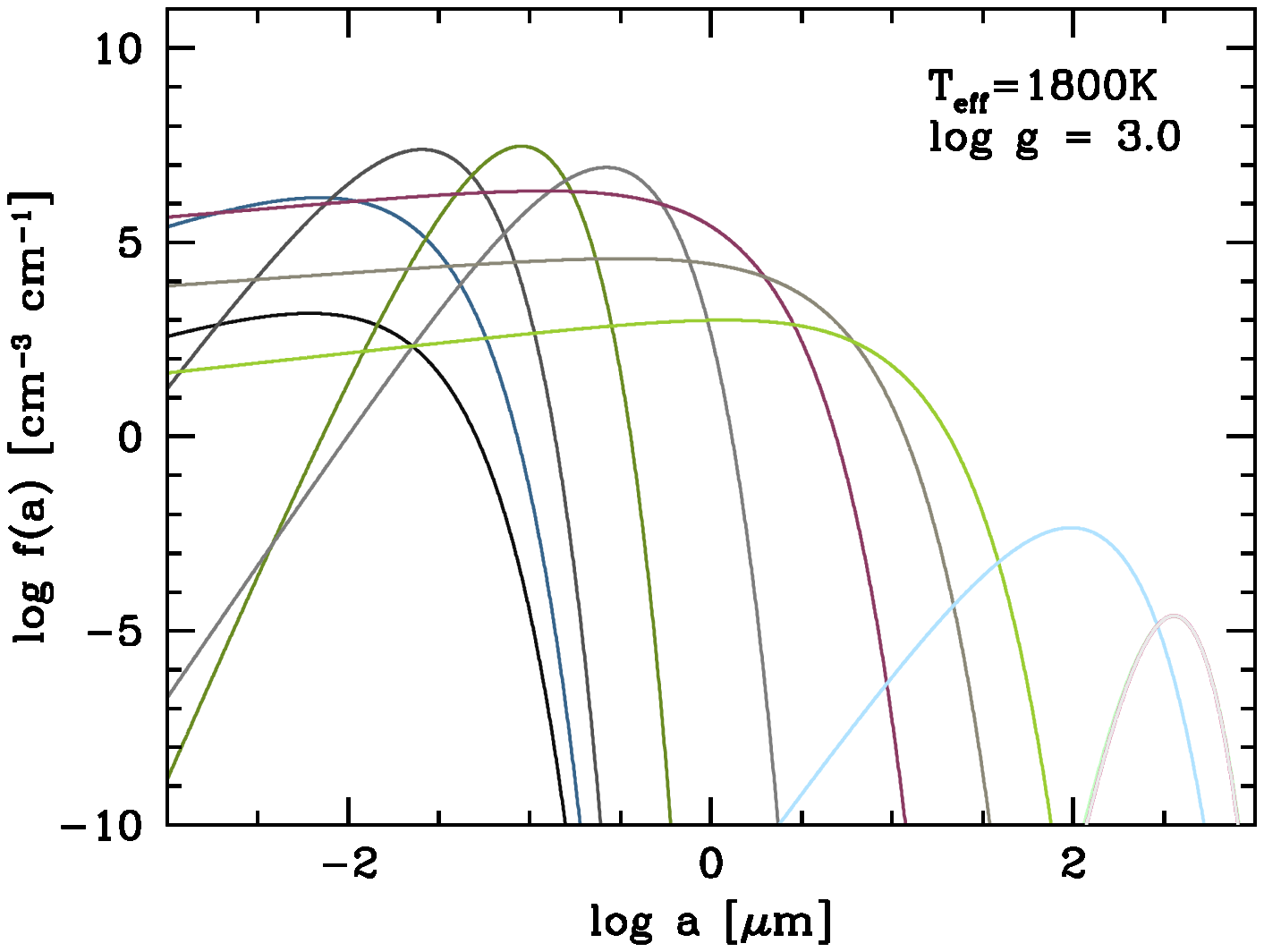}
  \includegraphics[width=8.5cm]{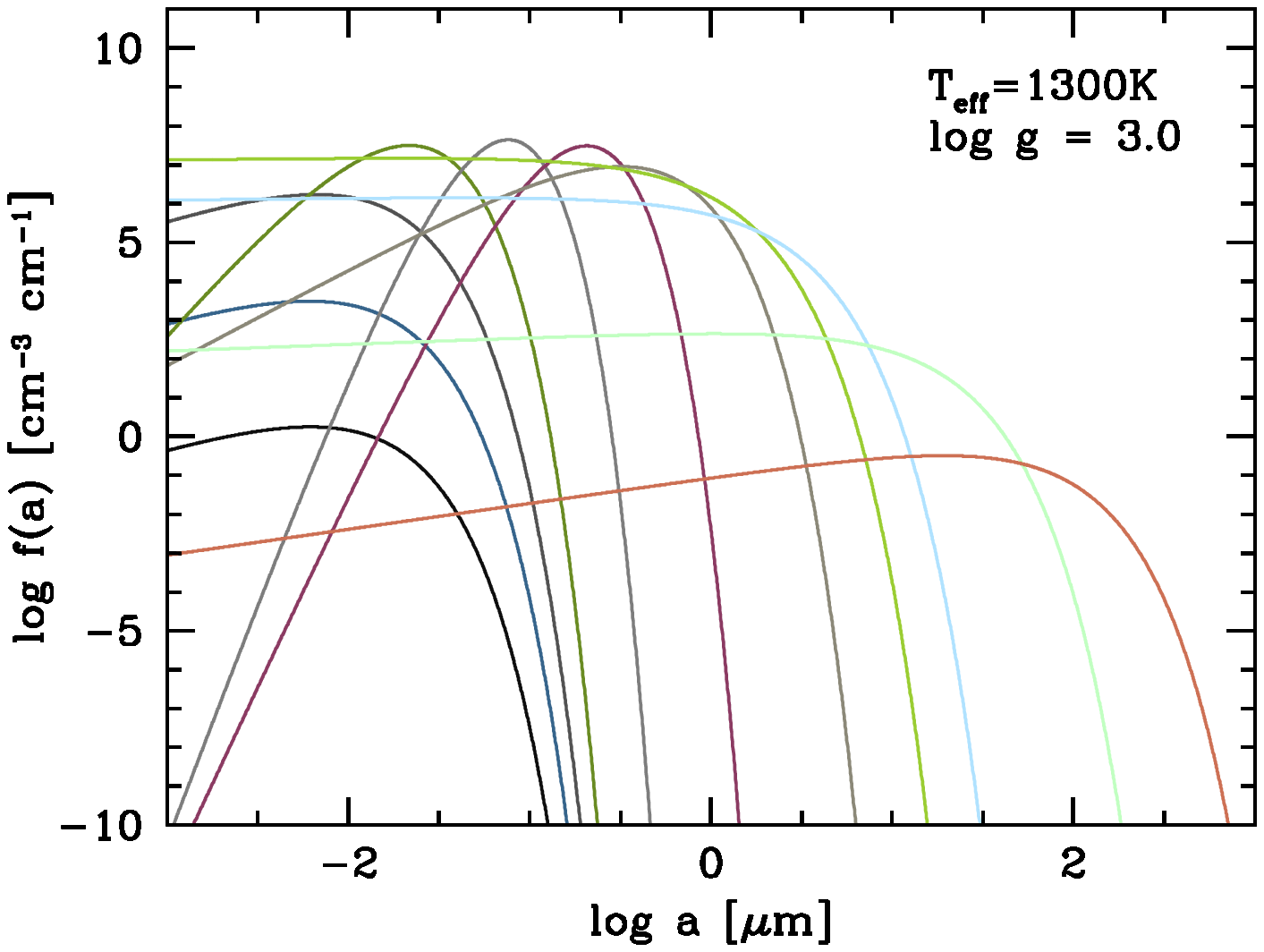}
  % --- Figure 8 ---     
  \caption{Grain size distribution function $f(a,z)$ [cm$^{-4}$] for
     a number of selected cloud altitudes. The broad distributions on the l.h.s. of all
     plots correspond to high altitudes. In the left upper figure, the
     deeper distributions are very close to a $\delta$-function and not
     presentable anymore.}
  \label{faz1800}
\end{figure*}

The calculation of the gas phase element abundances  allow for a
discussion of the metallicity in dust forming atmospheres (see
Fig.~\ref{metal1800}), \ie the logarithmic differences of the gas
abundances with respect to the solar values.  The metallicities are
not the same -- not even similar -- for different elements. The
strongest depletions occur for Ti, Mg and Si, for which the
metallicities reach values as low as $-6$ to $-7$. The depletion of S
and Ca is much less significant ($>\!-0.5$) in comparison.  Concerning
Ca, this is questionable because it is due to our limited choice of
solids. There is only one Ca bearing solid species considered, CaTi$_3$[s],
and since Ti is less abundant than Ca, Ca cannot be locked up
completely into grains.  It is important to note, however, that
according to our models, the gas depletion factors are {\it
kinetically} limited by a maximum factor of about $10^6$. Larger
depletion factors simply cannot occur because the depletion timescale
(see Paper V) would exceed the mixing timescale. In contrast, complete
phase equilibrium would imply that the depletion factors can reach
values as high as $10^{50}$ for low temperatures.
Figure~\ref{metal1800} also demonstrates that the metallicities change
with altitude and depend on the stellar parameter.

A special feature of our models is that the metal abundances in the
gas phase re-increase high above the cloud layer.  In our model, all
elements are constantly mixed upward, and although the mixing
timescale is as long as $\tau_{\rm mix}\!=\!10^{10}$s in the highest
layers, the timescale for a gaseous particle to find a surface to
condense on exceeds this timescale, because the dust abundance drops
steeply above the cloud layer. Consequently, the metal abundances
asymptotically re-approach the solar values high above the cloud
layer. The only exception is Ti, which can disappear from the gas
phase via nucleation.  This feature distinguishes our models from all
other published models, and possibly opens up a possibility to
discriminate between the models by detailed line profile observations
of neutral alkali and alkaline earth metal atoms. Our dust treatment
results in deeper and broader resonance lines of Na\,I, and K\,I in
the red part of the spectrum, which form high in the atmosphere
(Johnas et al. 2007).

\subsection{Grain size distribution}
\label{sec:gsd}

In Fig.~\ref{faz1800} we show the results for the potential
exponential grain size distribution function $f(a,z)\rm\,[cm^{-4}]$
(see Appendix~\ref{app:potexp})  with parameters derived from the
calculated dust moments $L_{\rm j}(z)$.  A number of  altitudes has been
 selected arbitrarily  throughout the entire dust cloud for visualisation.
 
The evolution of the grain size distribution through the cloud layer
are similar for models with the same $\log\,g$.  Considering the
$\log\,g\!=\!5$ models (upper row in Fig~\ref{faz1800}), the size
distribution starts relatively broad in the upper atmosphere where the
nucleation is active, because the constant creation and simultaneous
growth of the particles causes a broad distribution.  Once the
nucleation ceases, however, all particles merely shift in size space
by a constant offset $\Delta a$ due to further growth, which means a
narrowing in $\Delta(\log a)$. Since the particles grow by more than 4
orders of magnitude on their way down the atmosphere, the size
distribution finally becomes strongly peaked.

In the $\log\,g\!=\!3$ models (lower low in Fig~\ref{faz1800}),
the evolution of the size distribution function is more
complicated. As shown in Fig.~\ref{J*aquer}, the nucleation rate has a
a small shoulder as function of pressure on the left hand side. This
feature leads to a narrowing of $f(a)$ at first, followed by a re-widening of
the size distribution function with increasing depth. The nucleation
zone is closer to the convective zone and sometimes even overlaps with
the convective zone in the $\log\,g\!=\!3$ models. Therefore, although
the nucleation rate becomes tiny with increasing depth, it does not
vanish completely until about 1600\,K which still influences the size
distribution and keeps it broad.

\section{Simple radiative transfer modelling}
\label{sec:srt}

\subsection{Model description}
We simulate the grain absorption features in two brown dwarf
atmosphere cases ($T_{\mathrm{eff}}\!=\!1800$\,K, $\log\,g\!=\!5$ and
$T_{\mathrm{eff}}\!=\!1300$\,K, $\log\,g\!=\!5$) and one example for a
giant-gas planet atmosphere ($T_{\mathrm{eff}}\!=\!1300$\,K,
$\log\,g\!=\!3$) using a simple radiative transfer code. The code
considers the solid-state opacities similar to Helling et
al. (2006)\footnote{More complex methods applicable to hydrodynamic
environments are given in e.g. Helling \& J{\o}rgensen (1998) and
Woitke (2006).}. Grain opacities are calculated according to their size
distribution $f(a,z)$ [cm$^{-4}$] and their solid material volume
composition $\Vs(z)$ (Figs.~\ref{fig:VOLMASSFRAC},~\ref{faz1800})
using the effective medium theory according to Bruggeman (1935) to
calculate the effective optical constants and Mie theory for spherical
particles to calculate the extinction efficiencies $Q_{\rm
ext}(a,\lambda)$. The main difference between the previous code and
the present one is that the grains are here assumed to be distributed
according to the potential exponential size distribution function (see
Appendix~\ref{app:potexp})
%with parameters $\{A,B,C\}(z)$ 
determined from the calculated dust moments at height $z$, whereas in
the previous model a $\delta$-function, $f(a,z)=\rho
L_0(z)\,\delta\big(a-\langle a\rangle(z)\big)$ representing the mean
particle size $\langle a\rangle$, was assumed.  The present code
includes the 12 solid species discussed in the previous sections (see
Table~\ref{tab:dustopac} for the references for the optical
constants). The total grain extinction cross-section (see inner
integral in Eq.~\ref{eq:tau}) is computed using a Gauss-Legendre
quadrature integration over the entire grain size distribution
function at given atmospheric height.

\begin{figure*}
\vspace*{-0.8cm}
  % --- Figure 9 ---  
\begin{tabular}{ccc}
%\rotatebox{-90}{T$_{\rm eff}=1800\,$K, $\log$g$=5.0$, solar} & \rotatebox{90}{T$_{\rm eff}=1300\,$K, $\log$g$=5.0$, solar} & \rotatebox{90}{T$_{\rm eff}=1300\,$K, $\log$g$=3.0$, solar}\\
\includegraphics[width=6cm, angle=180]{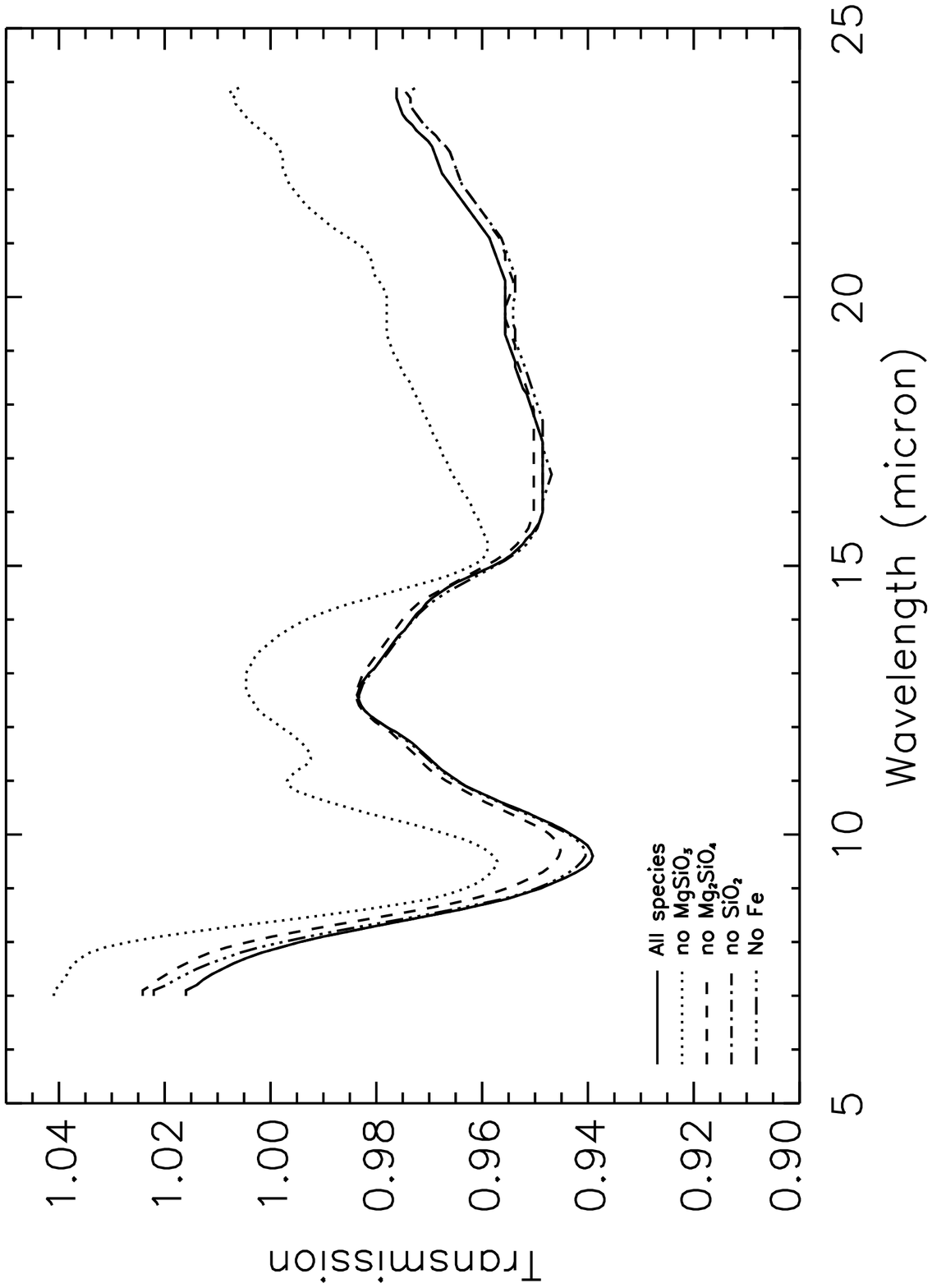}&
\hspace*{-0.7cm}\includegraphics[width=6cm, angle=180]{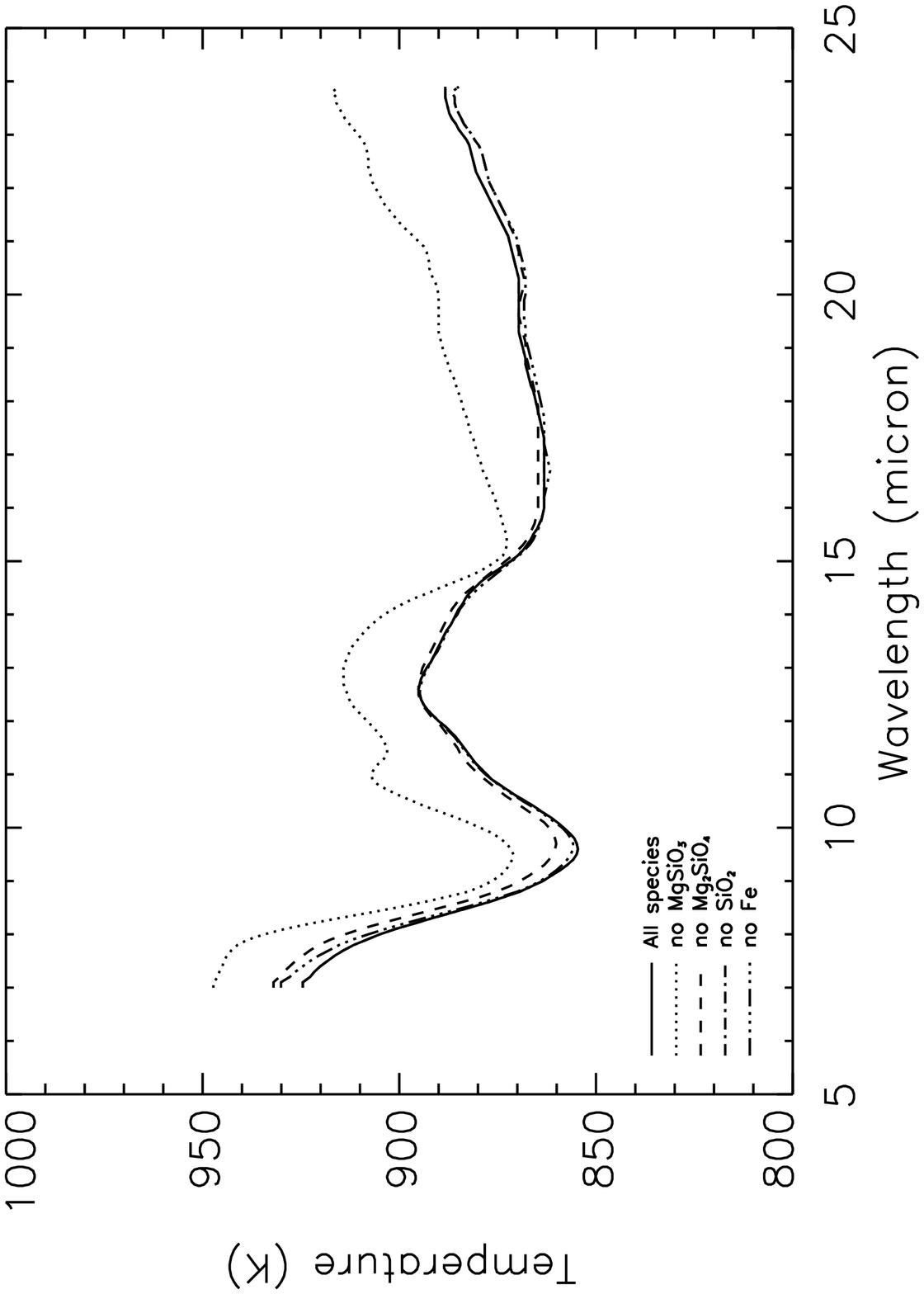}&
\hspace*{-0.7cm}\includegraphics[width=6cm, angle=180]{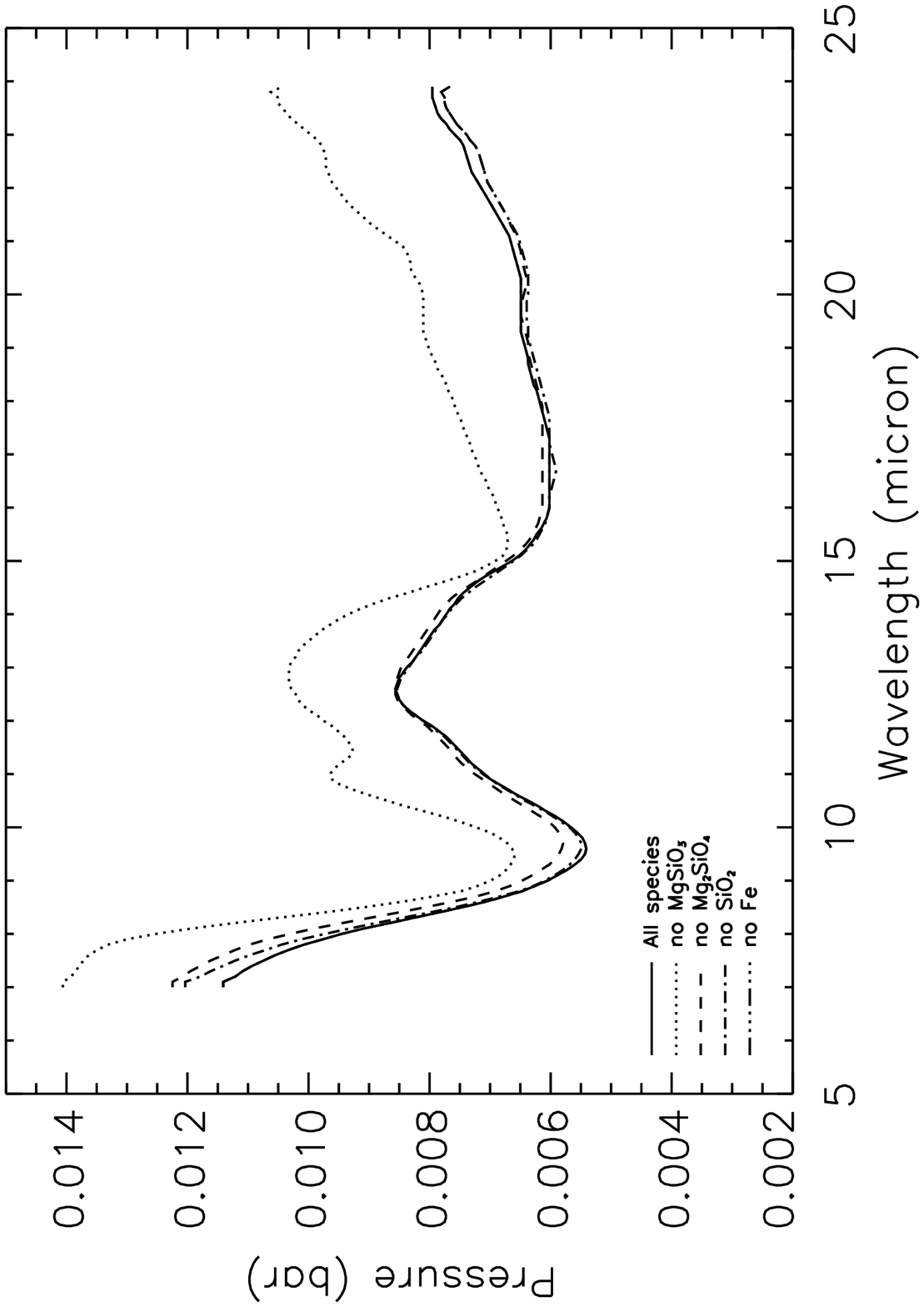}\\*[-0.5cm]
\includegraphics[width=6cm, angle=180]{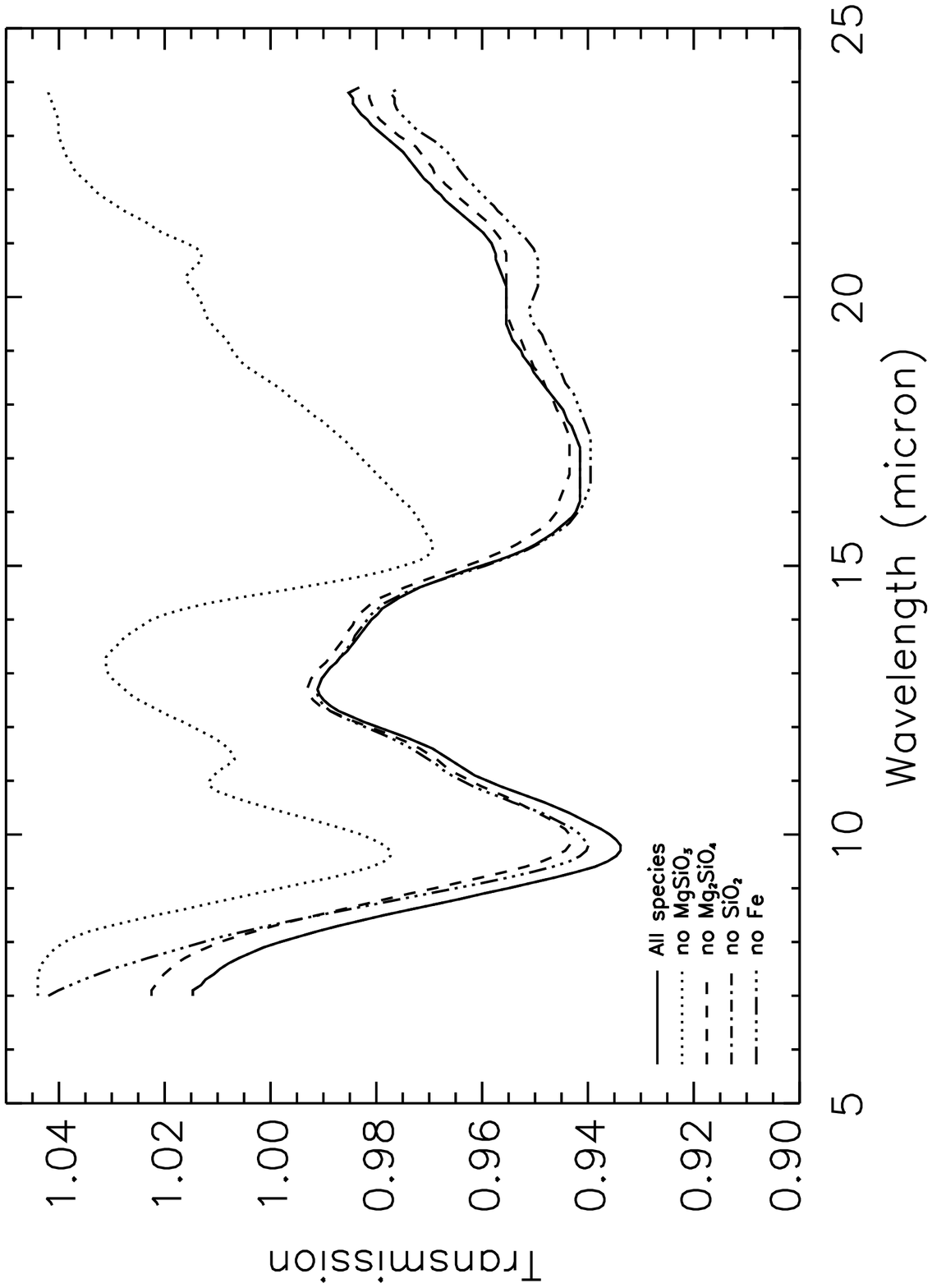}&
\hspace*{-0.7cm}\includegraphics[width=6cm, angle=180]{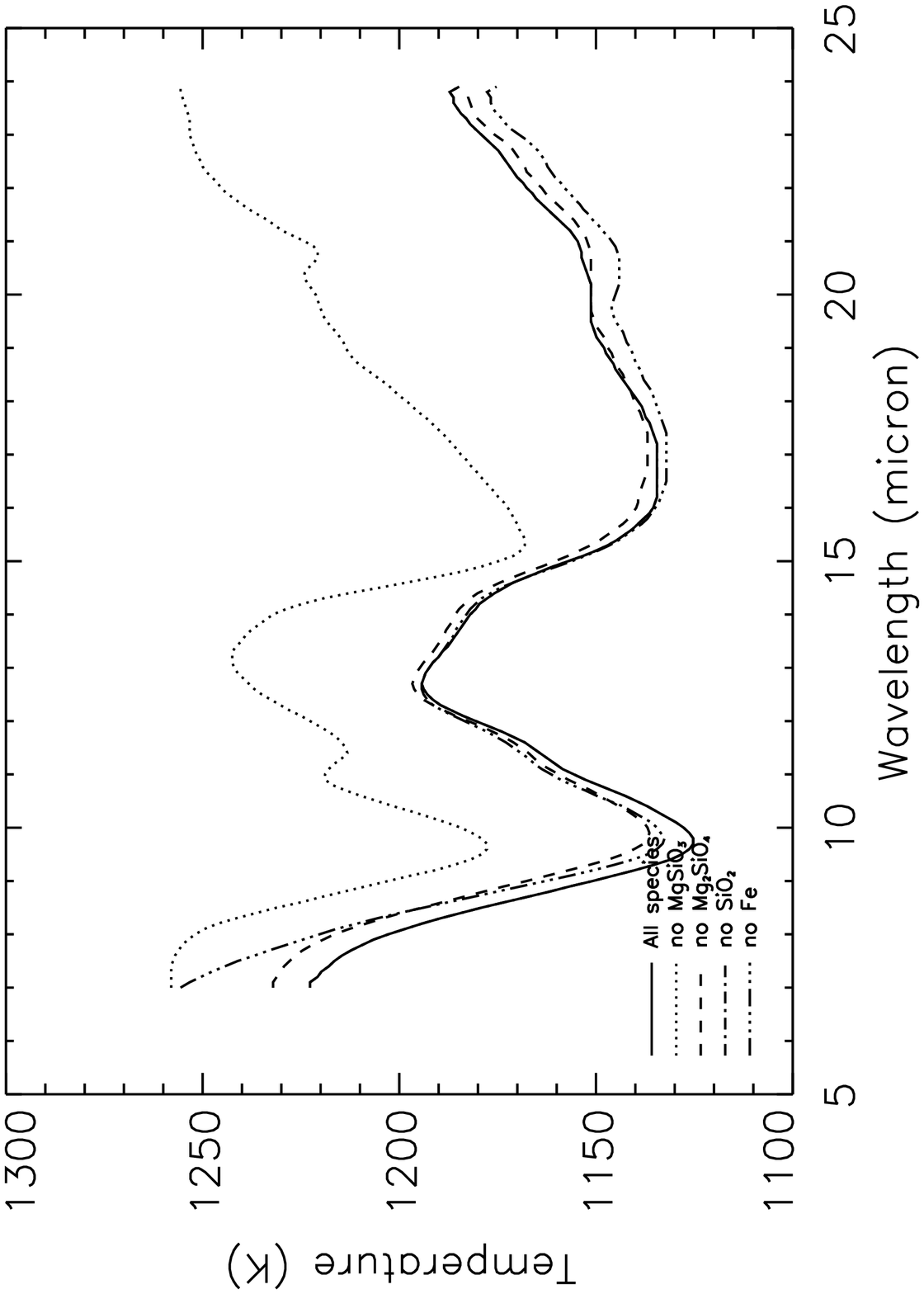}&
\hspace*{-0.7cm}\includegraphics[width=6cm, angle=180]{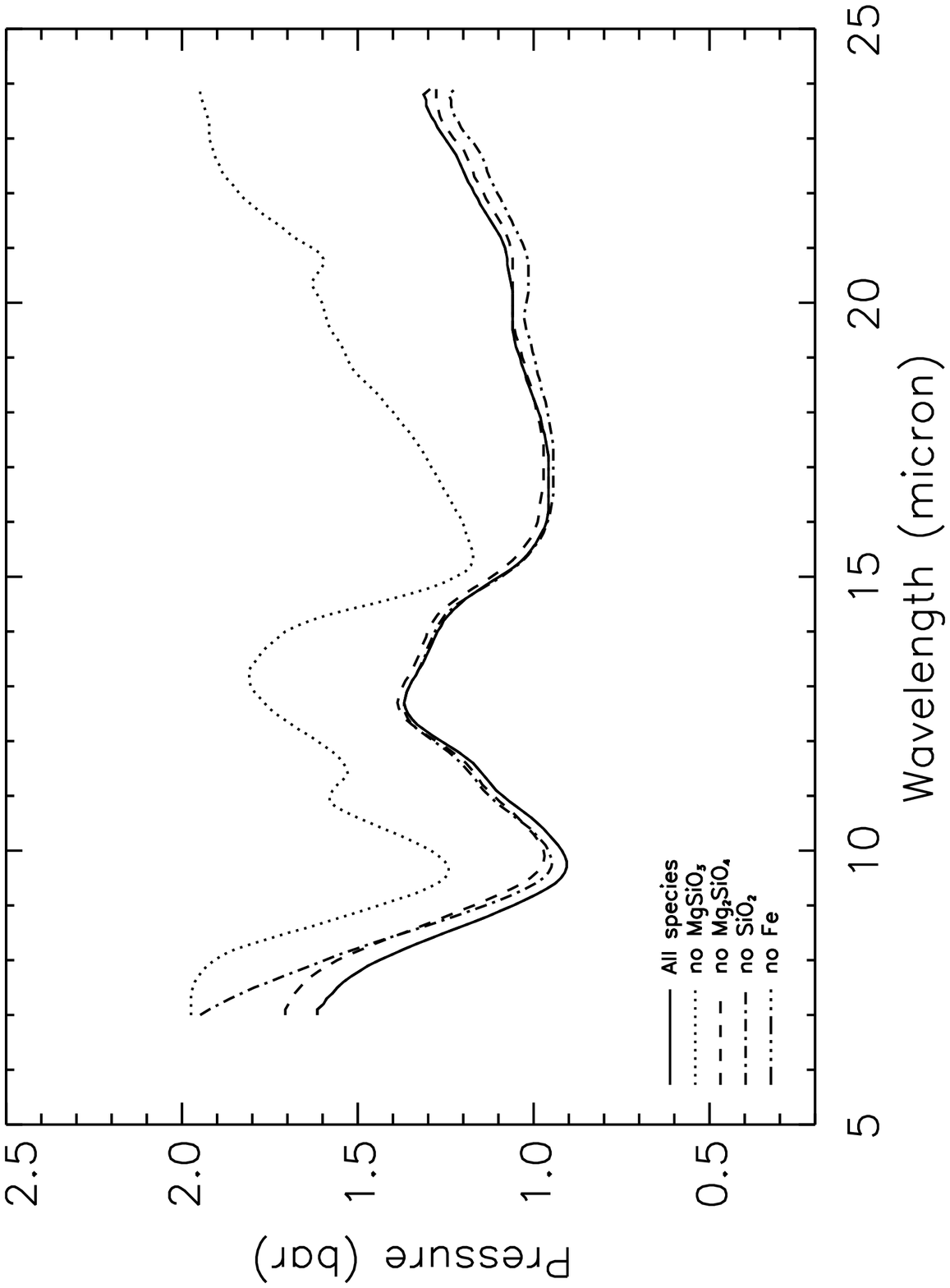}\\*[-0.7cm]
\includegraphics[width=6cm, angle=180]{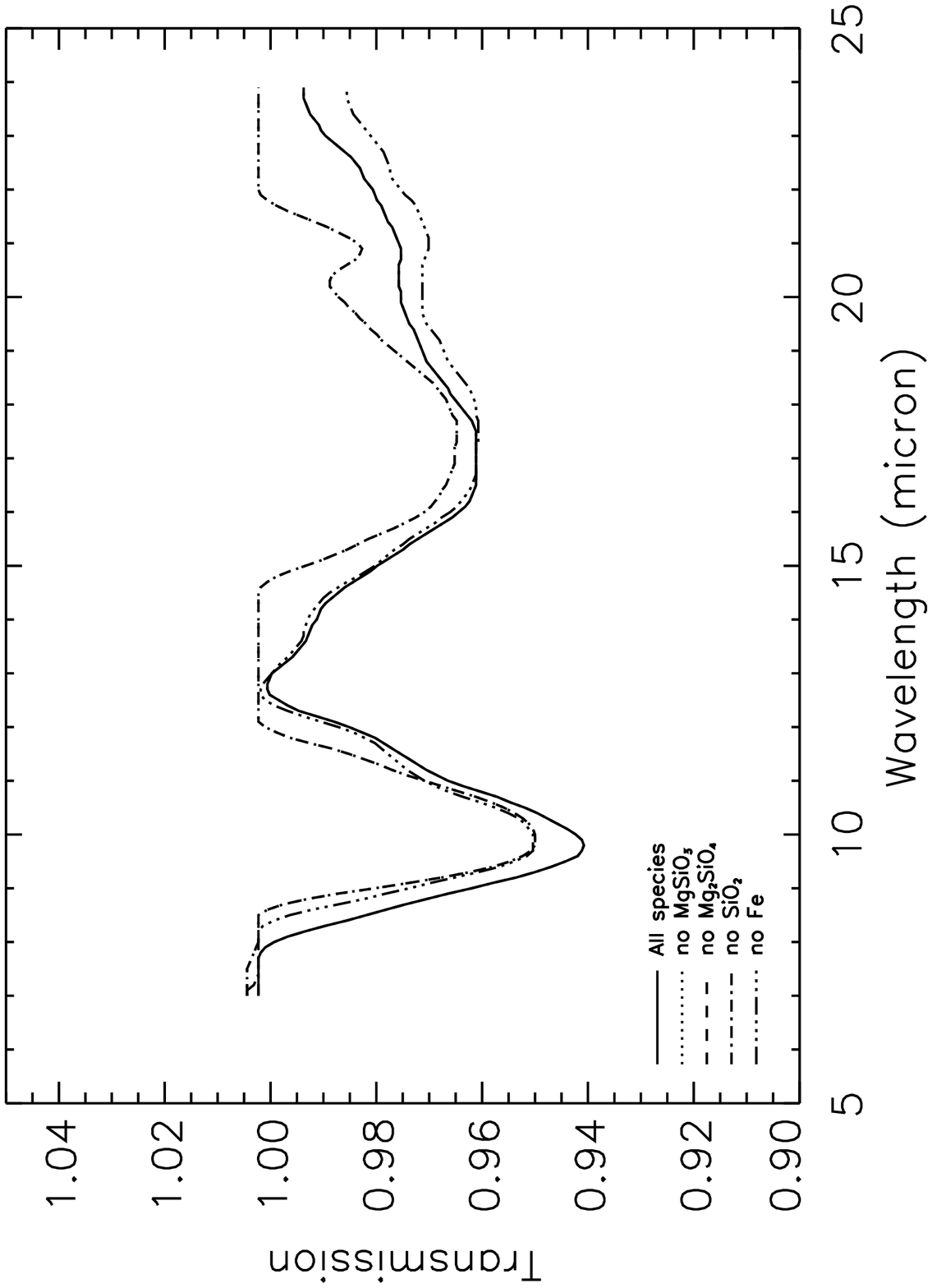}&
\hspace*{-0.7cm}\includegraphics[width=6cm, angle=180]{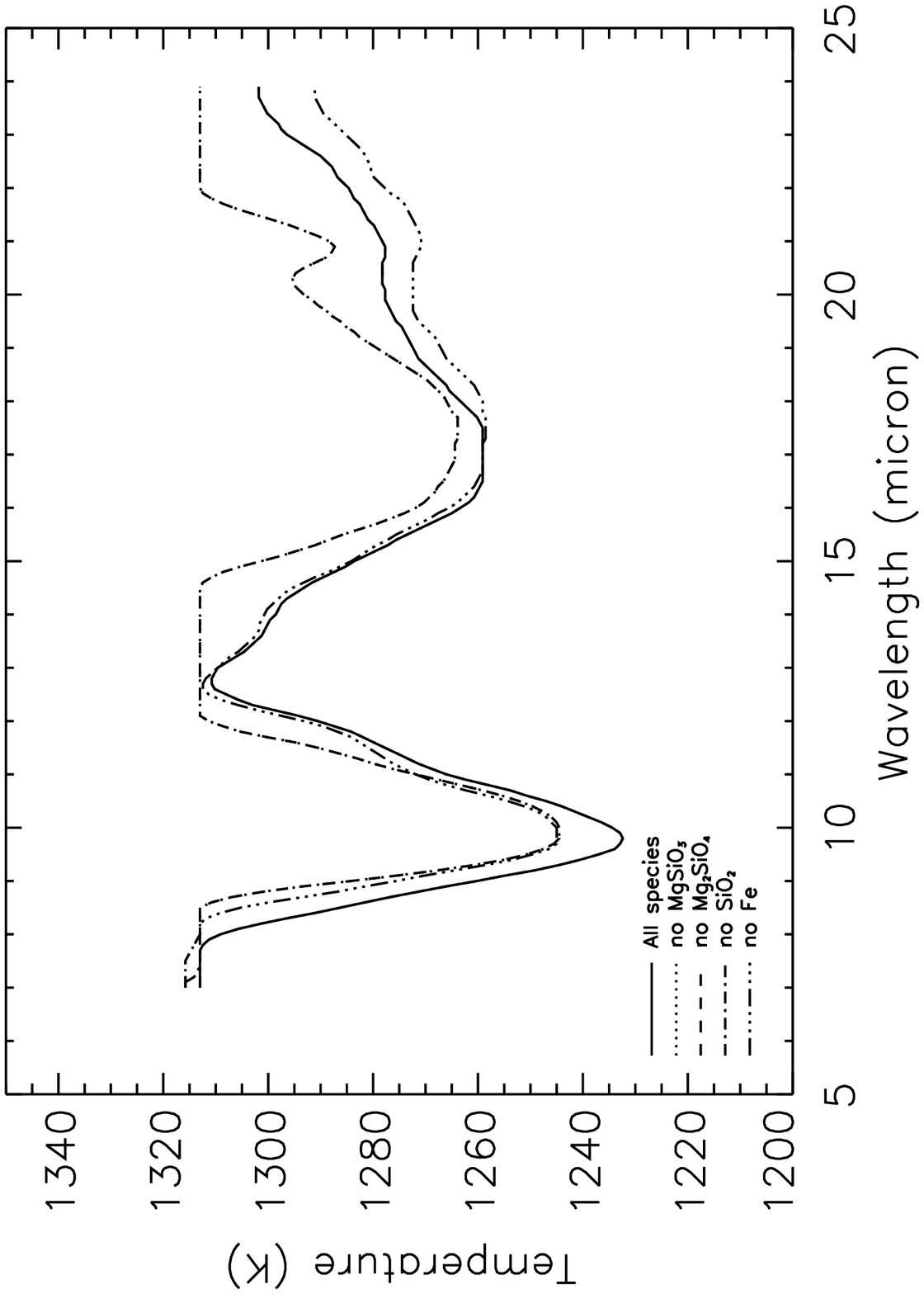}&
\hspace*{-0.7cm}\includegraphics[width=6cm, angle=180]{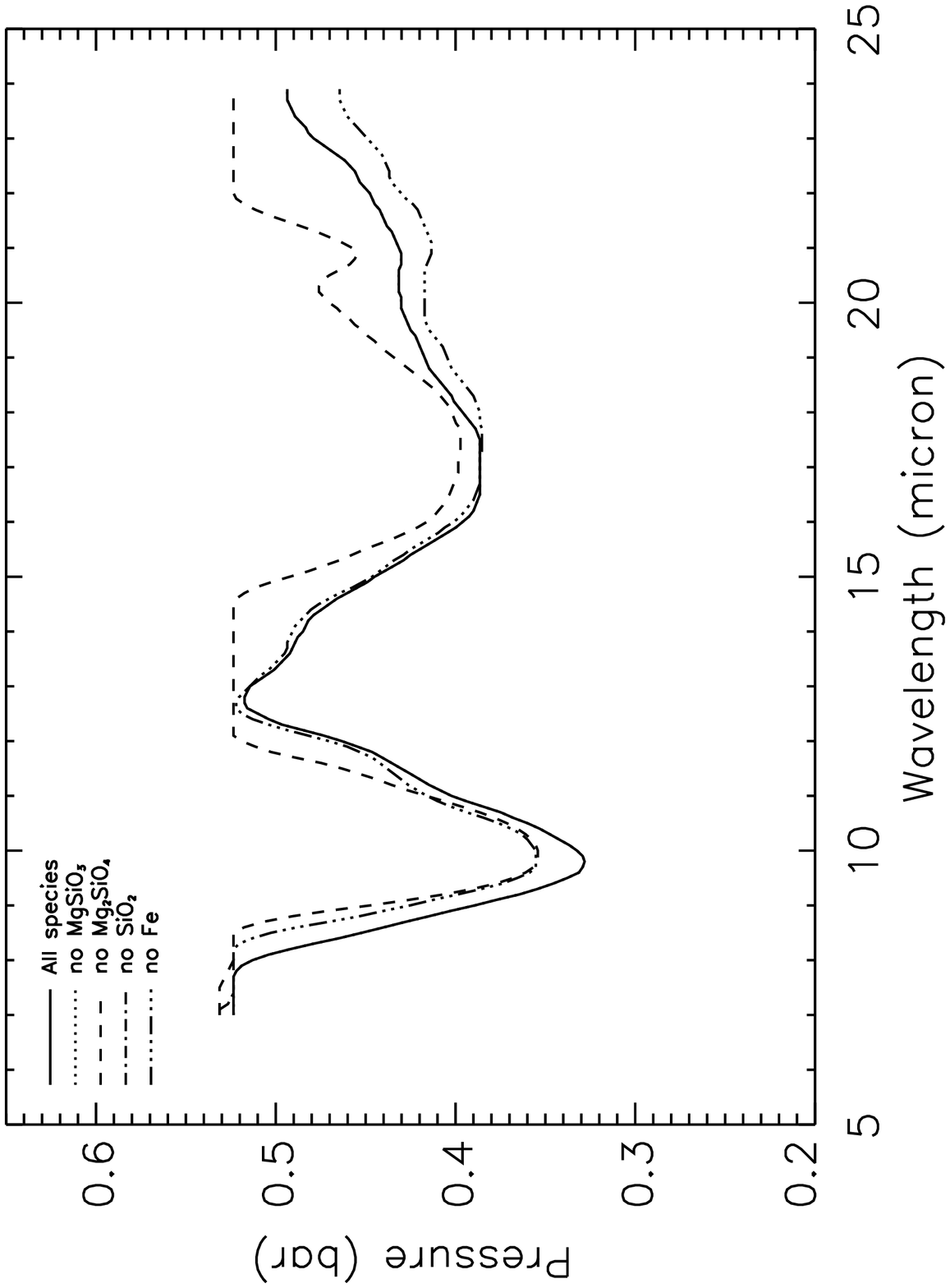}
\end{tabular}

  \caption{{\bf Upper row:} $T_{\rm eff}\!=\!1800$\,K, $\log\,g\!=\!5$;
	{\bf $2^{\rm nd}$ row:} $T_{\rm eff}\!=\!1300$\,K, $\log\,g\!=\!5$;
        {\bf Lower row:}        $T_{\rm eff}\!=\!1300$\,K, $\log\,g\!=\!3$. 
        {\bf Left column:} transmission spectra
	$F_\lambda\,\Big(T[z_0(\lambda)]\Big) \,\Big/B_\lambda(T_{\rm
	bb})$.
        {\bf Middle and right columns:} local temperature
        $T[z_0(\lambda)]$ and pressure $p[z_0(\lambda)]$,
        respectively, where optical depth reaches unity.  9.7 $\mu$m
        is the Si--O stretching mode of (SiO2, silicates), 16-19
        $\mu$m the bending mode.}
  \label{transm1800}
\end{figure*}

We  focus  the radiation transfer on the 7$\mu$m to 20$\mu$m wavelength
range, which encompasses the Si-O stretching mode (centred at 9.7
 $\mu$m) and Si-O bending mode (around 18  $\mu$m) of crystalline
silicates (Mg$_2$SiO$_4$ and MgSiO$_3$) and quartz (SiO$_2$). The
model calculates the optical depth of the dust component as 
\begin{equation}
  \tau^{\mathrm{dust}}_\lambda(z) = 
     \int\limits_0^z \int\limits_0^\infty
     f(a,z')\;\pi a^2 Q_{\rm ext}\Big(a,\lambda,\Vs(z')\Big)\;da\,dz'
  \label{eq:tau}
\end{equation}
 and determines the geometrical depth $z_0(\lambda)$ where
$\tau^{\mathrm{dust}}_\lambda(z_0)=1$ for each wavelength. Furthermore, 
we replace the opacity of the solid species by vacuum to estimate their
respective effects on the output spectrum.   To calculate the
transmission spectra (first row Fig.~\ref{transm1800}), we remove a
blackbody emission as continuum ($B_{\lambda}(T_{\rm bb}$)) from the spectrum ($F_{\lambda}$) as 
\begin{equation}
  \mbox{transmission} = F_\lambda\,\Big(T\big[z_0(\lambda)\big]\Big)
                       \,\Big/B_\lambda(T_{\rm bb})\ ,
\end{equation}
 \ie all transmission curves for one model atmosphere are divided 
by the same black body continuum: (T$_{\rm eff}$, $\log$\,g; T$_{\rm bb}$) =\\ \{(1800K, 5.0; 1310K), (1300K, 5.0; 1205K), (1300K, 3.0; 910K)\}.

\subsection{Results}
\label{sec:specs}

The absorption features can be distinguished from the continuum in all
cases. But they amount to a maximum of 6\% only at 9.7$\mu$m in all
three cases. The positions of the absorption features (9.7$\mu$m and 17--18
 $\mu$m) are typical of pure absorption by amorphous silicates, which
implies that scattering by the larger grains in the size distribution
does not contribute significantly to the total cross-sections. This
result differs from the single grain size situation where scattering
may modify the features, \eg by shifting the peak
absorptions.

The temperature and pressure at $z_0(\lambda)$, where
$\tau^{\mathrm{dust}}_\lambda=1$, are plotted below the transmission
spectra. The weak absorption reflects the shallow temperature and
pressure variations in the $\tau^{\mathrm{dust}}\!\approx\!1$
region. The contribution of quartz is negligible, as testified by the
same output spectrum whether quartz opacity (dash-dot line) is present
or not. The two silicate species $\rm MgSiO_3$ and $\rm Mg_2SiO_4$
contribute  about equally to the output spectrum (dotted
and dashed line).

Although grains exist even high in the atmosphere in this model,
their  number densities are too small to effect the opacity until
the region of rapid grain growth is reached.  The rapid grain growth
implies a concomitant sudden rise in opacities. The phenomenon is
particularly acute in the $(T_{\mathrm{eff}}\!=\!1800\,{\rm
K},\log\,g\!=\!5)$ model.  Therefore, the absorption features probe
 in particular the cloud deck.  The inclusion of gas opacities in
the radiative transfer model will complicate the 7--20 $\mu$m spectra,
definitively rendering the search for silicates features in brown
dwarf atmospheres difficult.

\begin{table}
\label{tab:dustopac}
\begin{center}
\caption{Reference for the optical constant of {\bf amorphous materials} used in the radiative
transfer modelling.}
\begin{tabular}{ll}
\hline
\hline
\noalign{\smallskip}
Solid species & Reference\\
\noalign{\smallskip}
\hline
\noalign{\smallskip}
TiO$_2$[s]     & Ribarsky et al. (1985)\\
SiO$_2$[s]     & Henning et al. (1997)\\
SiO[s]         & Philipp in Palik (1985)\\
Mg$_2$SiO$_4$[s] & J\"ager et al. (2003)\\
MgSiO$_3$[s]   & Dorschner et al. (1995) \\
MgO[s]         & Hofmeister et al. (2003)\\
Al$_2$O$_3$[s] & Begemann et al. (1997)\\
Fe[s]          & Ordal et al. (1985)\\
FeO[s]         & Henning et al. (1995)\\
FeS[s]         & Begemann et al (1994)\\
Fe$_2$O$_3$    & {\small http://www.astro.uni-jena.de/Laboratory/}\\
               & {\small OCDB/index.html}\\ 
                 %Amaury Triaud (Private communication)\\               
CaTiO$_3$[s]   & Posch et al. (2003)\\
\noalign{\smallskip}
\hline
\end{tabular}
\end{center}
\end{table}

\section{Conclusive summary}

Based on a detailed micro-physical description of the formation of
composite dust particles including nucleation, growth/evaporation and
drift we have  investigated the formation, chemical composition
and spectral appearance of quasi-static cloud layers in quasi-static
brown dwarf and gas-giant atmospheres. Our models show that
\begin{itemize}

\item  There is one  single cloud layer which stretches from about
$(1000\!-\!1300)\,$K to $(1600\!-\!1800)\,$K, which shifts 
to slightly higher temperatures for higher $\log\,g$ values.

\item The material composition of the cloud particles changes
continuously through the cloud layer. The seed particles created
high in the atmosphere are simultaneously covered by all kinds of solid
materials, favouring the abundant condensates like MgSiO$_3$[s],
Mg$_2$SiO$_4$[s], SiO$_2$[s] and Fe[s]. As the particles sink in
deeper, the higher temperatures enforce partial evaporation which
stepwise purifies the grains, leaving only the most stable condensates
like Fe[s] and Al$_2$O$_3$[s] at cloud base.

\item The material composition described above  is primarily a
function of local temperature and robust against changes in
$T_{\rm eff}$ and $\log\,g$.

\item  The mean sizes of the cloud particles increase continuously
through the cloud layer. The particles are found to be very small in
the high atmospheric layers ($\langle a\rangle\!\approx\!0.01\,\mu$m) with a
relatively broad size distribution, but as large as several
$100\,\mu$m with a strongly peaked size distribution at cloud base,
where they finally shrink due to evaporation and dissolve into the
surrounding gas. The maximum particle sizes reached are larger for
smaller $\log\,g$.

\item The gas is highly supersaturated in the upper atmosphere, 
where the nucleation takes place, but reaches a state close to
saturation from the cloud deck downward concerning the abundant
medium-temperature condensates. For high altitudes in general and for
high-temperature condensates in particular, phase equilibrium is not
valid, since the depletion timescale exceeds the mixing timescale.

\item  No cloud layer will form without the formation of seed
particles, which is different from terrestrial planets where seed
particles can be swapped up from the crust.

\item The consequence of the nucleation process is that the seed
forming elements are stronger depleted from the gas phase in the high
atmosphere than those elements which need an alien surface to
condense on, which leads to a highly Ti-depleted upper atmosphere
in our model. This finding could be useful to identify first
nucleation species in brown dwarfs and gas-giants.

\item Our models predict a much smaller degree of gas phase depletion
as compared to phase-equilibrium models.  The maximum metallicity
depression is about 6 orders of magnitude in our model for Mg, Si, Fe
and Al. Therefore, age estimates based on phase-equilibrium models
may have overestimated the brown dwarf's age considerably.

\item As a result of the phase-non-equilibrium, the carriers of
the metallic resonance lines, e.g. Na\,I, K\,I,   remain more abundant at
high altitudes which should make these lines deeper and broader in the
red part of the spectrum. This has by now been shown in Johnas et al. (2008).

\item The oxygen-depletion is not large enough to turn the
pre-mordially oxygen-rich substellar atmosphere into a carbon-rich
atmosphere with a carbon-to-oxygen ratio $>1$.

\item The broad dust absorption features attributed to the Si-O
stretching mode (centred at 9.7\,$\mu$m) and the Si-O bending mode
(around 18\,$\mu$m) should be present in brown dwarf spectra, but
their maximum absorption is weak ($<\!6\%$) and hence difficult to
detect.  The spectra primarily probe the cloud deck between about
$850\,$K and $1300\,$K, where Mg$_2$SiO$_4$[s] and MgSiO$_3$[s] are the
most important solid species. The positions of the dust absorption
features (9.7$\,\mu$m and 17--18$\,\mu$m) are typical of pure absorption
by amorphous silicates, which implies that scattering by the larger
grains in the size distribution does not contribute significantly to
the total cross-sections.

\end{itemize}

%\begin{acknowledgements}
%\end{acknowledgements}

\begin{appendix}
\section{Grain size distribution function}
\label{app:sizedist}

In order to find the dust particle size distribution function 
$f(a)\,\rm[cm^{-4}]$ we switch to another set of dust moments $K_j$ where the
integrals are performed in radius-space $a$ (Dominik et al. 1986, Gauger
et al. 1990) rather than in volume space $V$ (Dominik et
al. 1993). Substitution for $V\!=\!(4\pi a^3)/3$ yields
\begin{equation}
    K_j = \int_{a_\ell}^{\infty}\!\!\!f(a)\,a^j\,da 
      \,=\, \bigg(\frac{3}{4\pi}\bigg)^{j/3}\!\rho L_j 
  \label{eq:Kj}
\end{equation}
As argued in the main text, we are left with four known 
properties of the size distribution function, namely $K_1$, $K_2$,
$K_3$ and $K_4$. We will introduce a suitable functional  formula
for $f(a)$ with a set of  four free coefficients and will
determine these coefficients from the known dust moments. This exercise 
is carried out for two functions in the following.

\subsection{Double delta-peaked size distribution function}
\label{app:gsdf}

One option is to consider the superposition of two Dirac-functions
\begin{equation}
  f(a) = N_1\,\delta(a-a_1) + N_2\,\delta(a-a_2) \ ,
  \label{eq:parafunc1}
\end{equation}
where $\delta$ is the Dirac-function, $N_1, N_2\,\rm[cm^{-3}]$ are two
dust particle densities and $a_1, a_2\,\rm[cm]$ are two particle
radii. 

Assuming $a_1\!>\!a_\ell$ and $a_2\!>\!a_\ell$, where $a_\ell$ is the
lower integration size corresponding to $V_\ell$, the moments are
given by $K_j\!=\!N_1 a_1^{\,j} + N_2 a_2^{\,j}$, from which we can
determine the four free coefficients. $a_1$ is found to be the
positive root of
\begin{equation}
   a_1^2\,(K_2^2 - K_1 K_3) 
 + a_1\,(K_1 K_4 - K_2 K_3) 
        + (K_3^2 - K_2 K_4) \,=\,0
\end{equation}
and the other free coefficients are given by
\begin{eqnarray}
   a_2 &=& \frac{a_1 K_2 - K_3}{a_1 K_1 - K_2} \\
   N_1 &=& \frac{\big(a_1 K_1 - K_2\big)^3}
                {\big(a_1 K_2 - K_3\big)
                 \big(K_3 - 2a_1 K_2 + a_1^2 K_1\big)} \\
   N_2 &=& \frac{K_1 K_3 - K_2^2}
                {a_1\big(K_3 - 2a_1 K_2 + a_1^2 K_1\big)}
\end{eqnarray}

\subsection{Potential exponential size distribution function}
\label{app:potexp}

Another, more continuous option is to consider the function
\begin{equation}
  f(a) = a^B\!\exp\big(A\!-\!C\,a\big) \ ,
  \label{eq:parafunc2}
\end{equation}
which, for positive coefficients $A$, $B$ and $C$, is strictly
positive with a maximum at $B/C$.
Since there are only three coefficients in Eq.\,(\ref{eq:parafunc2}) 
we will determine them from $K_1$, $K_2$ and $K_3$.

To find a simple analytical solution, we extend the integration in
Eq.\,(\ref{eq:Kj}) to the interval $[-\infty\,...+\!\infty]$ which
usually introduces only a small error, since $f(a)$ is quickly
vanishing for small $a$ (see Fig.\,8). The result is
$K_j\!=\!\exp(A+\ln\Gamma(B\!+\!1\!+\!j)-(B\!+\!1\!+\!j)\ln C)$, by which
the free coefficients can be deduced:
\begin{eqnarray}
  B &=& \frac{2 K_1 K_3 - 3 K_2^2}{K_2^2 -K_1 K_3} \\
  C &=& (2\!+\!B)\,\frac{K_1}{K_2} \\
  A &=& \ln K_1 + (2\!+\!B)\ln C - \ln\Gamma(2\!+\!B) \,
\end{eqnarray}
where $\Gamma$ is the generalised factorial function with
$\Gamma(n\!+\!1) = n!$  for $n\in\varmathbb{N}_0$ and
$\Gamma(x\!+\!1) = x\,\Gamma(x)$ for $x\in\varmathbb{R}^+$.

\end{appendix}

\end{document}